\pdfoutput=1

\documentclass[11pt,twoside,a4paper,cmspaper,final,collab]{cms-tdr}
\def\svnVersion{490813P}\def\cmsCernNoTag{CERN-EP-2018-248}\def\cmsCernDate{\today}\def\cmsMessage{Published in the Journal of High Energy Physics as \href{http://dx.doi.org/10.1007/JHEP02(2019)074}{\doi{10.1007/JHEP02(2019)074.}}}

\begin{document}\cmsNoteHeader{EXO-16-053}

\hyphenation{had-ron-i-za-tion}
\hyphenation{cal-or-i-me-ter}
\hyphenation{de-vices}
\RCS$Revision: 489532 $
\RCS$HeadURL: svn+ssh://svn.cern.ch/reps/tdr2/papers/EXO-16-053/trunk/EXO-16-053.tex $
\RCS$Id: EXO-16-053.tex 489532 2019-02-20 14:43:34Z gomber $

\newlength\cmsTabSkip\setlength{\cmsTabSkip}{1ex}
\providecommand{\CL}{CL\xspace}

\newcommand{\currentL}{35.9}
\newcommand{\mpl}{\ensuremath{M_{\mathrm{Pl}}}}
\newcommand{\mD}{\ensuremath{M_\mathrm{D}}}
\newcommand{\mdm}{\ensuremath{m_{\text{DM}}}}
\newcommand{\mmed}{\ensuremath{M_{\text{med}}}}
\newcommand{\gDM}{\ensuremath{\text{\usefont{OT1}{cmr}{m}{it}g}_{\text{DM}}}}
\newcommand{\gq}{\ensuremath{\text{\usefont{OT1}{cmr}{m}{it}g}_{\cPq}}}
\newcommand{\ns}{\ensuremath{\,\text{ns}}\xspace}
\newcommand{\met}{\ptmiss}
\newcommand{\dR}{\ensuremath{\Delta R}}
\newcommand{\zinvg}{\ensuremath{\PZ(\to\PGn\PAGn){+}\Pgg}}
\newcommand{\zllg}{\ensuremath{\PZ(\to\ell\overline{\ell}){+}\Pgg}}
\newcommand{\wlng}{\ensuremath{\PW(\to\ell\PGn){+}\Pgg}}
\newcommand{\vg}{\ensuremath{\mathrm{V}{+}\Pgg}}
\newcommand{\gj}{\ensuremath{\Pgg{+}\text{jets}}}
\newcommand{\ttg}{\mbox{\ttbar\hspace{-0.3em}\Pgg}\xspace}
\newcommand{\ETg}{\ensuremath{E_{\mathrm{T}}^{\Pgg}}\xspace}
\newcommand{\sigetaeta}{\ensuremath{\sigma_{\eta\eta}}}
\newcommand{\DphiMETg}{\ensuremath{\Delta \phi(\ptmiss, \Pgg)}}
\newcommand{\minDphiMETj}{\ensuremath{\mathrm{min}\Delta \phi(\ptvecmiss, \ptvec^{\kern1pt\text{jet}})}\xspace}
\newcommand{\NZg}[1][]{\ensuremath{N^{\PZ\Pgg}_{#1}}}
\newcommand{\NWg}[1][]{\ensuremath{N^{\PW\Pgg}_{#1}}}
\newcommand{\RZll}[1][]{\ensuremath{R^{\PZ\Pgg}_{\ell\ell\Pgg#1}}}
\newcommand{\RZee}[1][]{\ensuremath{R^{\PZ\Pgg}_{\Pe\Pe\Pgg#1}}}
\newcommand{\RZmm}[1][]{\ensuremath{R^{\PZ\Pgg}_{\Pgm\Pgm\Pgg#1}}}
\newcommand{\RWl}[1][]{\ensuremath{R^{\PW\Pgg}_{\ell\Pgg#1}}}
\newcommand{\RWe}[1][]{\ensuremath{R^{\PW\Pgg}_{\Pe\Pgg#1}}}
\newcommand{\RWm}[1][]{\ensuremath{R^{\PW\Pgg}_{\Pgm\Pgg#1}}}
\newcommand{\fZW}[1][]{\ensuremath{f^{\PZ\Pgg}_{\PW\Pgg#1}}}
\newcommand{\Tll}[1][]{\ensuremath{T_{\ell\ell\Pgg#1}}}
\newcommand{\Tl}[1][]{\ensuremath{T_{\ell\Pgg#1}}}
\newcommand{\TK}[1][]{\ensuremath{T_{K#1}}}
\newcommand{\nhalo}[1][]{\ensuremath{n^{\text{halo}}_{K#1}}}
\newcommand{\NNPDFthree}{NNPDF3.0\xspace}
\newcommand{\NNPDFthreeone}{NNPDF3.1\xspace}
\newcommand{\DYRes}{\textsc{DYRes}\xspace}
\newcommand{\kappaSudakov}{\ensuremath{\kappa^{\text{EW Sudakov}}}}
\newcommand{\kappaPhoton}{\ensuremath{\kappa^{\text{EW }\Pq\Pgg}}}

\cmsNoteHeader{EXO-16-053}
\title{Search for new physics in final states with a single photon and missing transverse momentum in proton--proton collisions at $\sqrt{s} = 13\TeV$}

\date{\today}

\abstract{
A search is conducted for new physics in final states containing a photon and missing transverse
momentum in proton--proton collisions at $\sqrt{s} = 13\TeV$, using the data collected in 2016 by the CMS
experiment at the LHC, corresponding to an integrated luminosity of \currentL\fbinv. No
deviations from the predictions of the standard model are observed. The results are
interpreted in the context of dark matter production and models containing extra spatial
dimensions, and limits on new physics parameters are calculated at 95\% confidence level. For the two simplified
dark matter production models considered, the observed (expected) lower limits on the mediator
masses are both 950\,(1150)\GeV for 1\GeV dark matter mass.
For an effective electroweak--dark matter contact interaction, the observed (expected) lower limit on the
suppression parameter $\Lambda$ is 850\,(950)\GeV. Values of the effective Planck scale up
to 2.85--2.90\TeV are excluded for between 3 and 6 extra spatial dimensions.
}

\hypersetup{%
pdfauthor={CMS Collaboration},%
pdftitle={Search for new physics in final states with a single photon and missing transverse momentum in proton--proton collisions at sqrt(s) = 13 TeV},%
pdfsubject={CMS},%
pdfkeywords={CMS, physics, dark matter, gravitons}}

\maketitle

\section{Introduction}
\label{sec:introduction}

Production of events with a photon with large transverse momentum (\pt) and large missing transverse
momentum (\ptmiss) at the CERN LHC is a sensitive probe of physics beyond the standard model
(SM). This final state is often referred to as the ``monophoton'' signature, and has the advantage of
being identifiable with high efficiency and purity. Among the extensions of the SM that can be
studied with this final state are particle dark matter (DM) and large extra spatial dimensions.

\begin{figure}[hbtp]
  \centering
    \includegraphics[width=0.28\linewidth]{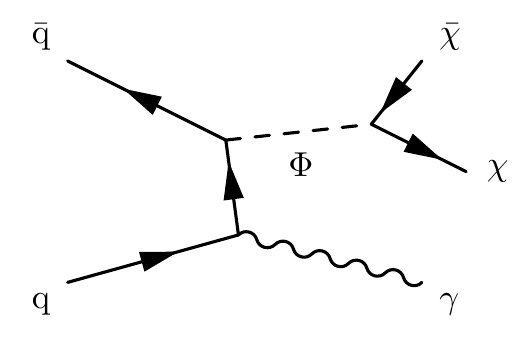}
    \includegraphics[width=0.28\linewidth]{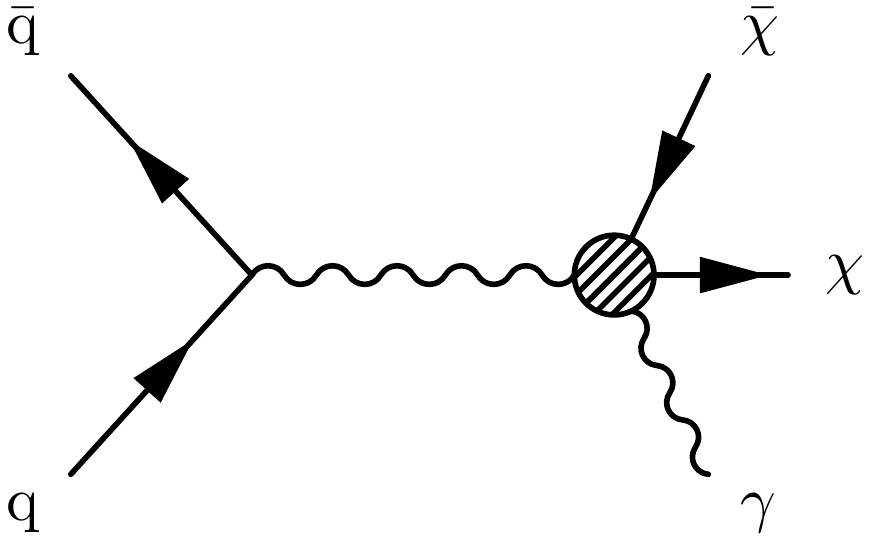}
    \includegraphics[width=0.28\linewidth]{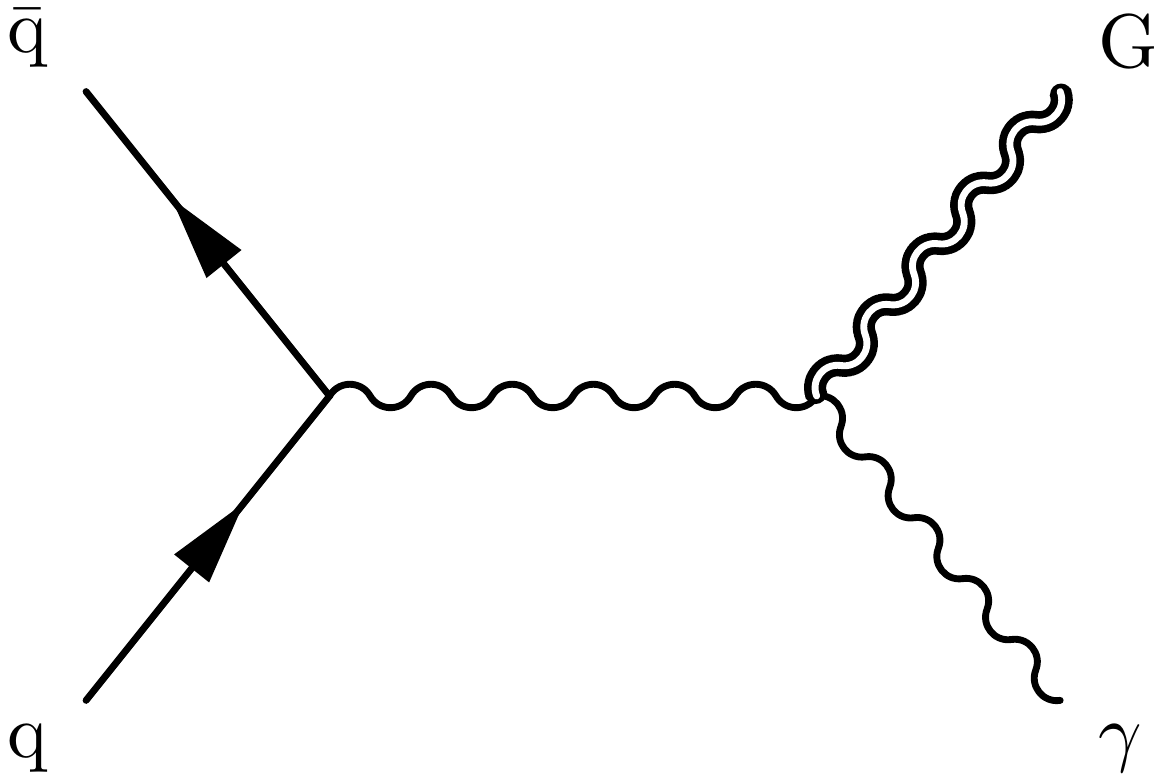}
    \caption{
      Leading order diagrams of the simplified DM model (left), EWK--DM effective interaction (center), and graviton (G) production in the ADD model (right), with a final state of a photon and large \ptmiss. Particles $\chi$ and $\overline{\chi}$ are the dark matter and its antiparticle, and $\Phi$ in the simplified DM model represents a vector or axial-vector mediator.
    }
    \label{fig:diagrams}
\end{figure}

At the LHC, the DM particles may be produced in high-energy proton--proton (\Pp\Pp) collisions, if they interact with the SM quarks or gluons via new couplings at the electroweak (EWK)
scale~\cite{DMCollid1,DMCollid2,DMLHC1}. Although DM particles cannot be directly detected, their production
could be inferred from the observation of events with a large \pt imbalance, when
high-energy SM particles recoil against the DM particle candidate. In DM production through a vector or axial-vector
mediator, a photon can be radiated from the incident quarks (Fig.~\ref{fig:diagrams}, left), resulting
in a monophoton final state. In the simplified models considered in
this analysis, Dirac DM particles couple to a vector or axial-vector mediator, which in turn couples
to the SM quarks. These models have been identified by the ATLAS--CMS Dark Matter
Forum~\cite{dmforum} as benchmarks to compare DM production sensitivity from various final
states. They are characterized by a set of four parameters: the DM mass \mdm, the
mediator mass \mmed, the universal mediator coupling to quarks \gq, and the mediator coupling
to DM particles \gDM. In this analysis, we fix the values of \gq\ and
\gDM\ to 0.25 and 1.0, respectively, and scan the \mmed--\mdm\ plane as recommended by the LHC Dark Matter Working Group~\cite{Boveia:2016mrp}.

It is also possible that the DM sector couples preferentially to the
EWK sector, leading to an effective interaction
$\PQq\PAQq\to\PZ/\Pgg^*\to\Pgg\chi\overline{\chi}$~\cite{Nelson:2013pqa}, where $\chi$ is the DM particle
(Fig.~\ref{fig:diagrams}, center). This model is characterized by a set of four parameters: the DM mass \mdm, the suppression
scale $\Lambda$, and the couplings $k_1$, $k_2$ to the $U(1)$ and $SU(2)$ gauge sectors, respectively.
In this analysis, we fix the values of $k_1$ and $k_2$ to 1.0, and set limits on $\Lambda$ at various
values of \mdm.

The model of large extra dimensions proposed by Arkani-Hamed, Dimopoulos, and Dvali (ADD)~\cite{ADD,ADD1}
postulates $n$ extra spatial dimensions compactified at a characteristic scale $R$ that reflects an
effective Planck scale \mD\ through $\mpl^2 \approx \mD^{n+2}R^n$, where
\mpl\ is the conventional Planck scale.  If \mD\ is of the same order as the
EWK scale ($M_\mathrm{EWK} \sim 10^{2}$\GeV), the large value of \mpl\ can be
interpreted as being a consequence of large-volume (${\sim}R^{n}$) enhancement from extra dimensional
space.  This model predicts a process $\Pq\Paq\to \Pgg\cPG$ (Fig.~\ref{fig:diagrams}
right), where \cPG\ represents one or more Kaluza--Klein gravitons, each of which can have any mass up to \mD.
Since the gravitons escape detection, this process leads to the monophoton final state.

In this paper we describe a search for an excess of monophoton events over the SM prediction.
Data collected by the CMS experiment in 2016, corresponding to an integrated luminosity of \currentL\fbinv,
are analyzed. Results are interpreted in the context of the three processes represented in Fig.~\ref{fig:diagrams}.

The primary irreducible background for the $\Pgg+\ptmiss$ signal is the SM \PZ boson production
associated with a photon, \zinvg. Other SM background processes include \wlng\ (where the charged lepton $\ell$ escapes detection),
$\PW\to\ell\Pgn$ (where $\ell$ is misidentified as a
photon), \gj, quantum chromodynamic (QCD) multijet events (with a jet misidentified as a photon), \Pgg\Pgg, t\Pgg, \ttg,
VV\Pgg\ (where V refers to a \PW\ or a \PZ boson), and \zllg. Additionally, a small residual
number of events from noncollision sources, such as beam halo~\cite{Beamhalo} interactions and
detector noise~\cite{Spike}, contribute to the total background.

A similar search in $\Pp\Pp$ collisions at $\sqrt{s} = 13\TeV$, based on a data set
corresponding to an integrated luminosity of 36.1\fbinv, has been reported by the
ATLAS experiment~\cite{Aaboud:2017dor}. No significant excess over the SM prediction was observed.
For the DM simplified model, a lower limit of 1200\GeV for both the vector and axial-vector mediator mass was set for
low DM masses under the same assumption on the new-physics coupling values. For the EWK--DM
effective interaction, a lower limit for the suppression parameter of the coupling was set at
790\GeV.

The previous search in the same final state by the CMS experiment~\cite{Sirunyan:2017ewk} is based
on $\sqrt{s} = 13\TeV$ data corresponding to an integrated luminosity of 12.9\fbinv, which is a
subset of the data analyzed in this paper. In addition to benefiting from a larger sample size, the
new analysis achieves improved sensitivity by using a simultaneous fit to the distributions of the \pt of the photon (\ETg) in
various signal and control regions to estimate the signal contribution, rather than the ``cut-and-count'' method deployed previously.

The paper is organized as follows. The CMS detector apparatus is described in Section~\ref{sec:detector}, along with the
algorithm used to reconstruct particles in pp collision events within the detector.
Section~\ref{sec:event_selection} lists the requirements that events must pass in order to be selected for inclusion in the signal
and control regions. Section~\ref{sec:modeling} lists the Monte Carlo generators used to model various signal and background
processes, and Section~\ref{sec:bkgestimate} describes the methods used to estimate the expected background yields in the signal
and control regions. These yields are tabulated in Section~\ref{sec:results}, which also presents the limits obtained for
each new physics model. The overall results are summarized in Section~\ref{sec:summary}. Appendix~\ref{sec:app} gives a detailed description
of the higher order corrections applied to the predicted differential cross sections of the leading background processes.

\section{The CMS detector and event reconstruction}
\label{sec:detector}

The central feature of the CMS apparatus is a superconducting solenoid of 6\unit{m} internal
diameter, providing a magnetic field of 3.8\unit{T}. Within the solenoid volume are a silicon pixel
and strip tracker, a lead tungstate crystal electromagnetic calorimeter (ECAL), and a brass and
scintillator hadron calorimeter (HCAL), each composed of a barrel ($\abs{\eta}< 1.48$) and two
endcap ($1.48 < \abs{\eta} < 3.00$) sections, where $\eta$ is the pseudorapidity. The ECAL consists
of 75\,848 lead tungstate crystals, with 61\,200 in the barrel and 7324 in each of the two endcaps.
In the $\eta$--$\phi$ plane, HCAL cells in the barrel map on to $5{\times}5$ arrays of ECAL
crystals to form calorimeter towers projecting radially outwards from close to the nominal interaction point.
Forward calorimeters extend the $\eta$ coverage provided by the barrel and endcap detectors. Muons
are detected in gas-ionization chambers embedded in the steel flux-return yoke outside the solenoid.

Events of interest are selected using a two-level trigger system~\cite{Khachatryan:2016bia}. The
first level, composed of custom hardware processors, uses information from the calorimeters and
muon detectors to select events at a rate of around 100\unit{kHz} within a time interval of less
than 4\mus. The second level, known as the high-level trigger (HLT), consists of a farm of
processors running a version of the full event reconstruction software optimized for fast
processing, and reduces the event rate to less than 1\unit{kHz} before data storage.
A more detailed description of the CMS detector, together with a definition of the coordinate system
and kinematic variables, can be found in Ref.~\cite{Chatrchyan:2008aa}.

Global event reconstruction follows the particle-flow (PF) algorithm~\cite{CMS-PRF-14-001}, which
aims to reconstruct and identify each individual particle in an event with an optimized combination
of all subdetector information. In this process, the identification of the particle type (photon,
electron, muon, charged hadron, neutral hadron) plays an important role in the determination of the
particle direction and energy. Photons are identified as ECAL energy clusters not linked to the
extrapolation of any charged particle trajectory to the ECAL, while electrons are identified as ECAL
energy clusters with such a link. Muons are identified as tracks in the central tracker consistent
with either a track or several hits in the muon system, and associated with calorimeter deposits compatible
with the muon hypothesis. Charged hadrons are identified as tracks neither identified as
electrons nor as muons. Note that all three types of charged candidates can be associated to a
reconstructed interaction vertex through their tracks. Finally, neutral hadrons are identified as
HCAL energy clusters not linked to any charged hadron trajectory, or as ECAL and HCAL energy
excesses with respect to the expected charged hadron energy deposit.

Reconstruction of $\Pp\Pp$ interaction vertices proceeds from tracks using a deterministic annealing
filter algorithm~\cite{Chatrchyan:2014fea}. The reconstructed vertex with the largest value of
summed physics-object $\pt^2$ is taken to be the primary $\Pp\Pp$ interaction vertex. Here, the
physics objects are the jets, clustered using the jet finding
algorithm~\cite{Cacciari:2008gp,Cacciari:2011ma} with the tracks assigned to the vertex as inputs,
and the associated missing transverse momentum, taken as the negative vector sum of the \pt of those
jets. These definitions of the jets and missing transverse momentum are specific to the context of
vertex reconstruction, and are distinct from the definitions in the remainder of the analysis, as
described in the following.

For each event, hadronic jets are clustered from these reconstructed PF candidates using
the anti-\kt algorithm~\cite{Cacciari:2008gp,Cacciari:2011ma} with a distance parameter of 0.4. The jet momentum is determined as the
vector sum of all particle momenta in the jet. Because of the large number of additional
\Pp\Pp\ interactions within the same or nearby bunch crossings (pileup), particles emerging
from multiple interactions can be clustered into a jet. To mitigate this effect, charged candidates
associated with vertices other than the primary one are discarded from clustering, and an offset
correction is applied to the \pt of the jet to subtract the remaining contributions~\cite{Khachatryan:2016kdb}. Jet energy
corrections are derived from simulation to bring, on average, the measured response of jets to that of particle level
jets. Measurements on data of the momentum balance in dijet, $\text{photon} +
\text{jet}$, $\PZ + \text{jet}$, and multijet events are used to account for any residual
differences in jet energy scale in data and simulation. Additional selection criteria are applied to
each jet to remove jets potentially dominated by anomalous contributions from various subdetector
components or reconstruction failures~\cite{Khachatryan:2016kdb}.

The missing transverse momentum vector (\ptvecmiss) is defined as the negative vector sum of the
transverse momenta of all PF candidates in an event. The magnitude of \ptvecmiss\ is the missing
transverse momentum, \ptmiss.

ECAL clusters are identified starting from cluster seeds, which are ECAL crystals with energies above
a minimum threshold, that must also exceed the energies of their immediate neighbors.
Topological clusters are grown from seeds by adding adjacent crystals with energies above a lowered
threshold, which could include other seeds. A topological cluster is finally separated into distinct
clusters, one for each seed it contains, by fitting its energy distribution to a sum of Gaussian-distributed
contributions from each seed.

Photon and electron reconstruction begins with the identification of ECAL
clusters having little or no observed energy in the corresponding HCAL region.
For each candidate cluster, the reconstruction algorithm searches for hits in the pixel and strip trackers
that can be associated with the cluster. Such associated hits are called electron seeds, and are
used to initiate a special track reconstruction based on a Gaussian sum
filter~\cite{Adam:2003kg,Khachatryan:2015hwa} which is optimized for electron tracks. The energy of electrons
is determined from a combination of the electron momentum at the primary interaction vertex as determined by the tracker,
the energy of the corresponding ECAL cluster, and the energy sum of all bremsstrahlung photons spatially
compatible with originating from the electron track. An ECAL cluster with no associated electron seed, or with a significant
energy excess relative to any compatible tracks, gives rise to a photon candidate.
The energy of a photon is determined only from its corresponding ECAL cluster.

\section{Event selection}
\label{sec:event_selection}

The integrated luminosity of the analyzed data sample is $(\currentL\pm0.9)$\fbinv~\cite{CMS:2017sdi}. The
data sample is collected with a single-photon trigger that requires at least one photon candidate
with $\pt >165\GeV$. The photon candidate must have $H/E < 0.1$ to discriminate against jets, where
$H/E$ is the ratio of HCAL to ECAL energy deposits in the central calorimeter tower corresponding to the candidate.
The photon energy
reconstructed at the HLT is less precise relative to that derived later in the offline
reconstruction. Therefore, the thresholds in the trigger on both $H/E$ and \ETg, are less restrictive than their offline counterparts.  The trigger efficiency is
measured to be about 98\% for events passing the analysis selection with $\ETg >175\GeV$.

From the recorded data, events are selected by requiring $\ptmiss > 170\GeV$ and at least one photon
with $\ETg > 175\GeV$ in the fiducial region of the ECAL barrel ($\abs{\eta} < 1.44$). Photon
candidates are selected based on calorimetric information, isolation, and the absence of an
electron seed, where the first two categories of the selection requirements are designed to
discriminate the photon candidates from electromagnetic (EM) showers caused by hadrons, and the
third is designed to discriminate photon candidates from electrons.

The calorimetric requirements for photons comprise $H/E < 0.05$ and $\sigetaeta < 0.0102$. The variable $\sigetaeta$, described in
detail in Ref.~\cite{Khachatryan:2015iwa}, represents the width of the EM shower in the
$\eta$ direction, which is generally larger in showers from hadronic activity. For a photon
candidate to be considered as isolated, the scalar sums of the transverse momenta of charged hadrons,
neutral hadrons, and photons within a cone of $\dR = \sqrt{\smash[b]{(\Delta \eta)^2 + (\Delta \phi)^2}} < 0.3$
around the candidate photon must all fall below a set of corresponding bounds chosen to give 80\% signal
efficiency. Only the PF candidates
that do not overlap with the EM shower of the candidate photon
are included in the isolation sums.
Ideally, the isolation sum over PF charged hadrons should be computed using only the candidates
sharing an interaction vertex with the photon candidate. However, because photon candidates are not
reconstructed from tracks, their vertex association is ambiguous. When an incorrect vertex is
assigned, nonisolated photon candidates can appear isolated. To reduce the rate for
accepting nonisolated photon candidates, the maximum charged-hadron isolation value
over all vertex hypotheses (worst isolation) is used. The above criteria select efficiently both unconverted photons and photons 
undergoing conversion in the detector material in front of the ECAL.

Stray ECAL clusters produced by
mechanisms other than $\Pp\Pp$ collisions can be misidentified as photons. In particular, anomalous ECAL
energy deposits resulting from the interaction of particles in the ECAL photodetectors, from here
on referred to as ``ECAL spikes'', as well as  beam halo muons that accompany proton beams and
penetrate the detector longitudinally have been found to produce spurious photon candidates at
nonnegligible rates. The ECAL spike background is reduced by requiring that the photon candidate
cluster must comprise more than a single ECAL crystal. To reject the beam halo induced EM showers,
the ECAL signal in the seed crystal of the photon cluster is required to be within $\pm 3\ns$ of
the arrival time expected for particles originating from a collision. In addition, the maximum of
the total calorimeter energy summed along all possible paths of beam halo particles passing through the cluster
(halo total energy), calculated for each photon candidate, must be below 4.9\GeV. The two requirements
combined with the shower shape constraint suppress the beam halo background effectively, while retaining
95\% of signal photons. Furthermore, using features described in Section 5.4, the signal
region is split into two parts according to $\phi$ to constrain the beam halo normalization. The region defined by
$\abs{\sin(\phi)} < \sin(0.5)$ is called the horizontal region, and its complement in $\phi$ is
called the vertical region.

Events with a high-\pt photon and large \ptmiss are subjected to further requirements to suppress
SM background processes that feature a genuine high-energy photon, but not a significant amount of \ptmiss.
One such SM process is \gj, where an apparent large \ptmiss is often the result of a mismeasured jet energy.
In contrast to signal processes, \ptmiss is typically smaller than \ETg\ in these events,
so requiring the ratio of \ETg\ to \ptmiss to be less than 1.4 rejects this background effectively
with little effect on signal efficiency. Events are also rejected if the minimum opening angle
between \ptvecmiss and the directions of the four highest \pt jets, \minDphiMETj, is less than
0.5. Only jets with $\PT > 30\GeV$ and $\abs{\eta} < 5$ are considered in the
\minDphiMETj\ calculation. In the \gj\ process, rare pathological mismeasurement of \ETg\ can also
lead to large \ptmiss. For this reason, the candidate photon \ptvec and
\ptvecmiss must be separated by more than 0.5 radians. Another SM process to be rejected is \wlng,
for which events are vetoed if they contain an electron or a muon with $\PT > 10\GeV$ that is
separated from the photon by $\dR > 0.5$.

The residual contributions from the \wlng\ process, where the lepton could not be identified or was
out of the detector acceptance, are modeled by fitting to observed data, as described in Section~\ref{sec:bkgestimate}. The same method is
employed to model the contribution from the \zinvg\ process to the signal region. This method
utilizes control regions where one or two leptons (electrons or muons) are identified in addition to
the photon, as defined in the following.

The single-electron (single-muon) control region is defined by a requirement of exactly one electron (muon)
with $\PT > 30\GeV$ and $\abs{\eta} < 2.5\,(2.4)$ in addition to a photon requirement that is identical
to the one for the signal region. To suppress the contributions from large-\ptmiss processes
other than \wlng, the transverse mass
$\sqrt{\smash[b]{2\ptmiss\PT^{\ell}[1-\cos\Delta\phi(\ptvecmiss,\ptvec^{\ell})]}}$ must be less than $160\GeV$.
Additionally, for the single-electron control region, \ptmiss must be greater
than 50\GeV to limit the contribution from the \gj\ process, where a jet is misidentified as an
electron. Finally, the recoil vector $\vec{U} = \ptvecmiss + \ptvec^{\ell}$, which serves as this region's analogue for
\ptvecmiss in the signal region, must satisfy identical requirements to those for the \ptvecmiss in the signal region.

The dielectron (dimuon) control region is defined by exactly two electrons (muons) in addition to
the photon, with $60 < m_{\ell\ell} < 120\GeV$, where $m_{\ell\ell}$ is the invariant mass of the dilepton
system. The recoil vector of this region is $\vec{U} = \ptvecmiss + \sum\ptvec^{\ell}$ and must
satisfy identical requirements to those for the \ptvecmiss in the signal region.

\section{Signal and background modeling}
\label{sec:modeling}

Monte Carlo simulation is used to model the signal and some classes of SM background
events. For the leading order (LO) samples, the
\NNPDFthree~\cite{Ball:2014uwa} leading order (LO) parton distribution function (PDF) set is used with the strong coupling constant value $\alpha_{S}$ = 0.130, whereas for the next-to-leading-order (NLO) samples, the \NNPDFthreeone~\cite{Ball:2017nwa} next-to-next-to-leading-order (NNLO) PDF set with $\alpha_{S} = 0.118$ is employed. For the SM background processes, the primary hard interaction is simulated using the
\MGvATNLO\ version 2.2.2~\cite{Alwall:2014hca} generator at LO in QCD.
The simulated events for the \zinvg, \zllg, and \wlng\ background processes, collectively denoted as \vg, are generated with \MGvATNLO at LO in QCD with up to two extra partons
 in the matrix element calculations. These are then normalized to the NLO EW and NNLO QCD cross sections using correction factors described in Section 5.1. Parton showering
and hadronization are provided by \PYTHIA{} 8.212 with the underlying-event tune
CUETP8M1~\cite{tune}. Multiple simulated minimum bias events are overlaid on the primary interaction to model
the distribution of pileup in data. Generated particles are processed through the full
\GEANTfour-based simulation of the CMS detector~\cite{GEANT, GEANTdev}.

For the DM signal hypotheses, \MGvATNLO\ 2.2.2 is used to produce MC simulation samples at
NLO in QCD, requiring $\ETg > 130\GeV$ and $\abs{\eta^{\Pgg}} <2.5$. A large
number of DM simplified model samples are generated, with varying \mmed\ and \mdm. Similarly,
EWK--DM effective interaction samples are generated in a range of 1--1000\GeV
for the DM particle mass. For the ADD hypothesis, events are generated using \PYTHIA{} 8, requiring
$\ETg > 130\GeV$, with no restriction on the photon $\eta$. Samples are prepared in a grid of values for the
number of extra dimensions and \mD. The efficiency of the full event selection for these signal
models ranges between 0.06 and 0.29 for the DM simplified models, 0.44 and 0.46 for EW DM
production, and 0.23 and 0.30 for the ADD model, depending on the parameters of the models.

\section{Background estimation}
\label{sec:bkgestimate}

\subsection{\texorpdfstring{\zinvg}{Z (invisible) + gamma} and \texorpdfstring{\wlng}{W (leptonic) + gamma} background}
\label{subsec:vgamma}

The most significant SM background processes in this search are the associated production of a
high-energy \Pgg\ with either a \PZ\ boson that subsequently decays to a pair of neutrinos, or a \PW\ boson
that decays to a charged lepton and a neutrino. The two processes are denoted as \zinvg\ and
\wlng. Together, they account for approximately 70\% of the SM background, with 50\% from the former
and 20\% from the latter. Contributions from these two
background processes are estimated using observed data in the four mutually exclusive single-electron,
single-muon, dielectron, and dimuon control regions defined in Section~\ref{sec:event_selection}. The ratios
between the expected yields of these processes are constrained by MC simulations of \vg\ processes.

The individual MC simulation samples of \vg\ processes receive multiple correction factors.  First
is the selection efficiency correction factor $\rho$, which accounts for subtle differences
between simulation and observation in the reconstruction and identification efficiencies for various
particle candidates. The value of $\rho$ typically lies within a few percent of unity.  The second
factor is the higher-order QCD correction, which matches the distribution of the generator-level
\ETg\ to that calculated at NNLO in QCD using the \DYRes program~\cite{Catani:2015vma}. The third
factor further corrects the \ETg\ distributions to account for NLO EW effects, and is taken from
Refs.~\cite{Denner:2014bna,Denner:2015fca}, updated using the LUXqed17 PDF
set~\cite{Manohar:2017eqh}.

Four sources of systematic uncertainties considered for \ETg\ distribution ratios among the
\vg\ processes are PDFs, higher-order QCD corrections,
higher-order EWK corrections, and data-to-simulation correction factors $\rho$.
The PDF uncertainty is evaluated by varying the weight of each event using the weights provided
in the NNPDF set, and taking the standard deviation of the resulting \ETg\ distributions. This
uncertainty is considered fully correlated in the ratio between the \zinvg\ and \wlng\ processes,
\ie, the variation of the ratio is bounded by the ratios of the upward and downward variations.
Uncertainties related to higher-order QCD corrections are considered uncorrelated in the ratio
between the \zinvg\ and \wlng\ processes. Because EW corrections become increasingly important at higher
\ETg, but are known only up to NLO accuracy, their uncertainties are estimated by a special prescription
similar to that discussed in Ref.~\cite{Lindert:2017olm}, where independent degrees of freedom
are assigned to the uncertainty in the overall scale of the correction and the uncertainty in the
variation of the correlation with \ETg. Additionally, the full correction due to photon-induced
$\PZ+\Pgg$ and $\PW+\Pgg$ production cross sections is considered as an uncertainty.
Further details concerning the higher-order QCD and EWK corrections are given in Appendix~\ref{sec:app}.
Finally, data-to-simulation correction factors $\rho$ for the lepton identification efficiencies have associated
uncertainties that do not cancel when taking ratios between regions defined by different lepton selection requirements. The four uncertainties are all considered as correlated between the \ETg\ bins.

\begin{figure}[htbp]
  \centering
    \includegraphics[width=0.49\textwidth]{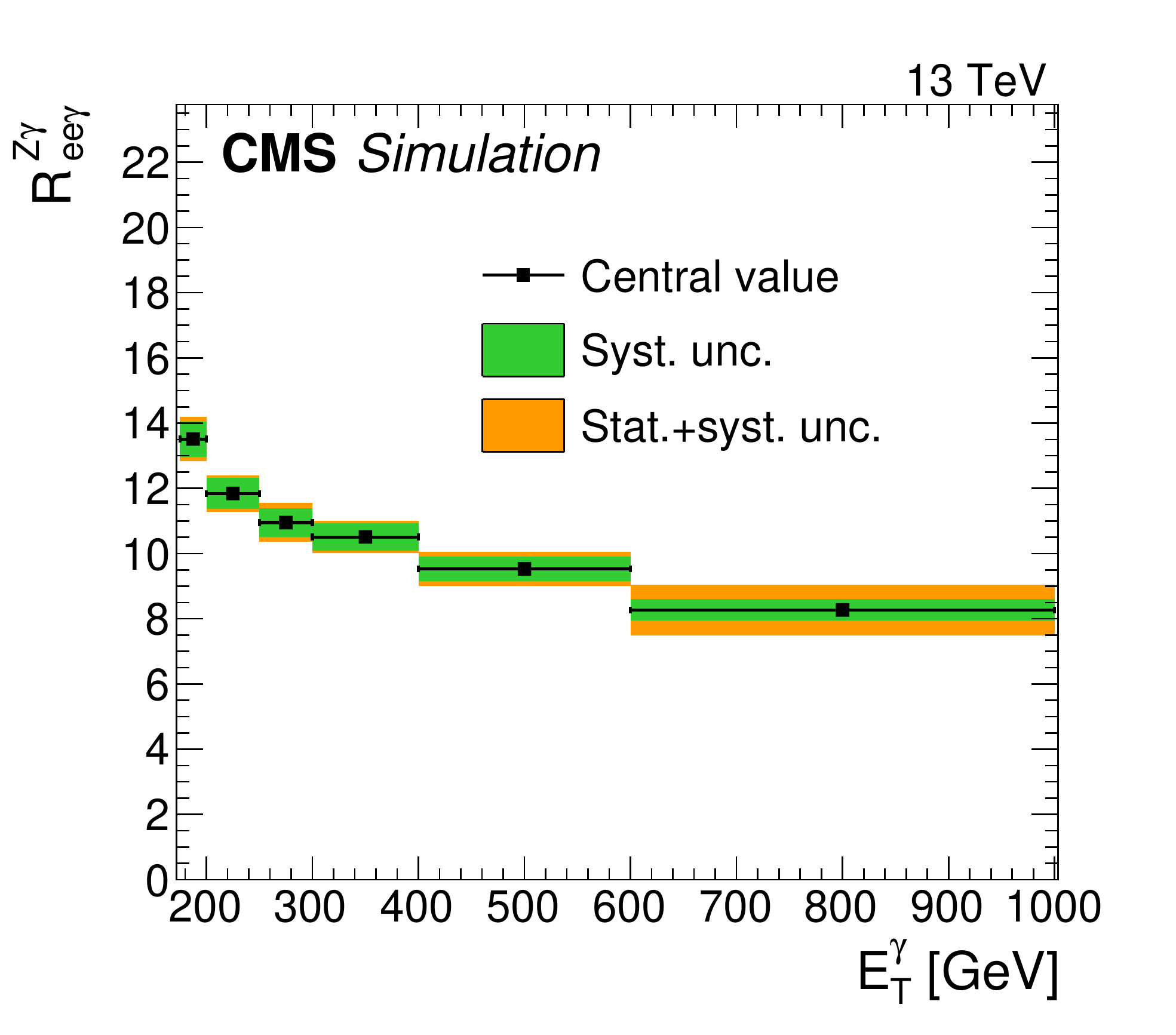}
    \includegraphics[width=0.49\textwidth]{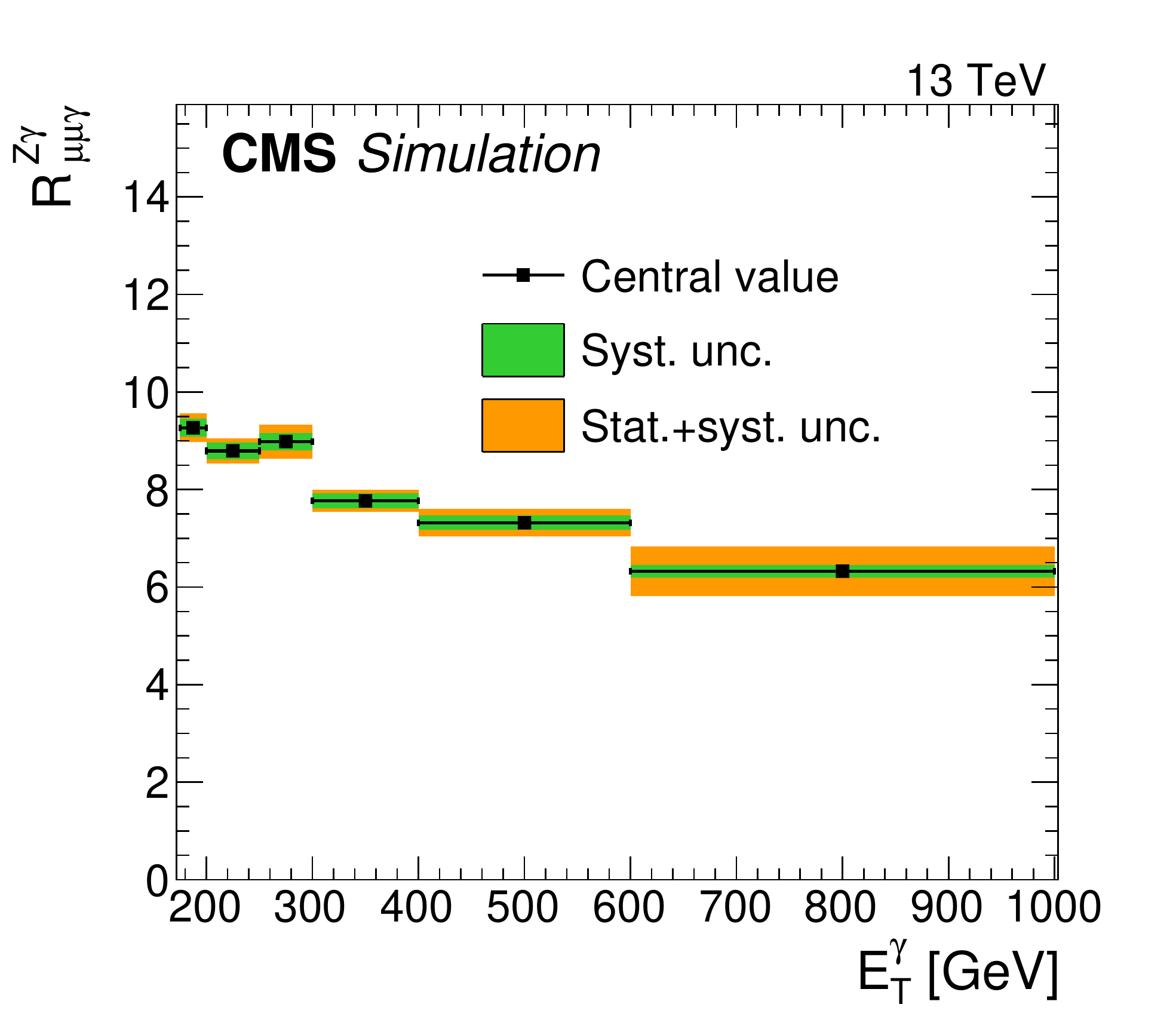}
    \caption{
      Transfer factors \RZee\ (left) and \RZmm\ (right). The uncertainty bands in green (inner) and orange (outer) show the systematic uncertainty, and the combination of systematic and statistical uncertainty arising from limited MC sample size, respectively. The systematic uncertainties considered are the uncertainties in the data-to-simulation correction factors $\rho$ for the lepton identification efficiencies. Simulated \zllg\ events are generated in two samples, one with generated \ETg\
required to be greater than 300 GeV, and one with a looser restriction. The \ETg\ bin centred at 270 GeV is close to the boundary between the two samples, where there are fewer generated events. The relatively large statistical fluctuation visible in the third bin of the right-hand figure results from this.
    }
    \label{fig:tf_z}
\end{figure}

The background estimation method exploits cancellation of some of the systematic uncertainties, both
experimental and theoretical, in the ratios of the photon \ETg\ distributions of \vg\ processes,
from here on referred to as ``transfer factors''. For example, in the transfer factor between the \zinvg\ and
\zllg\ processes, denoted \RZll, the uncertainties due to photon energy calibration,
jet energy resolution, and higher-order QCD effects are significantly reduced compared to when such effects
are considered for individual processes. The only uncertainties in the transfer factor
\RZll\ that do not largely cancel are those on lepton identification efficiency and the
statistical uncertainty due to the limited MC sample size. Figure~\ref{fig:tf_z} shows the transfer factor \RZee\ (\RZmm)
between the dielectron (dimuon) control region and the combined signal regions, for which the numerator
is the expected \zinvg\ yield in the combined signal regions and the denominator is the expected \zllg\ yield in the relevant control region.

Using the transfer factor \RZll, the total estimated event yield \Tll\ in each dilepton
control region in the $i^\mathrm{th}$ bin of the \ETg\ distribution can be expressed as
\begin{equation}
  \Tll[,i] = \frac{\NZg[i]}{\RZll[,i]} + b_{\ell\ell\Pgg,i},
\end{equation}
where \NZg\ is the number of \zinvg\ events in the combined signal regions and
$b_{\ell\ell\Pgg}$ is the predicted contribution from other background sources in the dilepton
control region, namely \ttg, VV\Pgg, and misidentified hadrons. The subscript $i$
indicates that the quantities are evaluated in bin $i$ of the \ETg distribution.

\begin{figure}[htbp]
  \centering
    \includegraphics[width=0.49\textwidth]{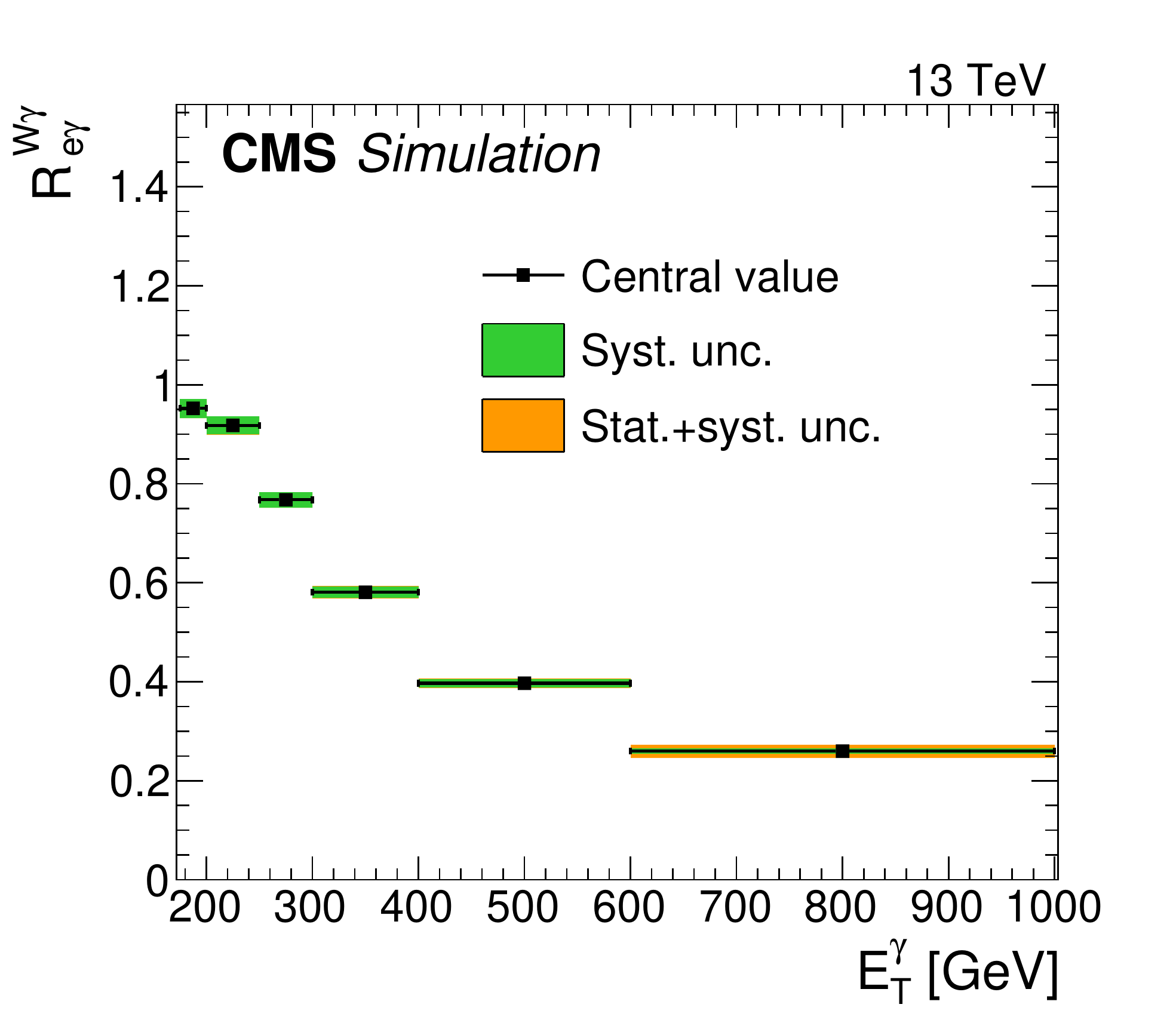}
    \includegraphics[width=0.49\textwidth]{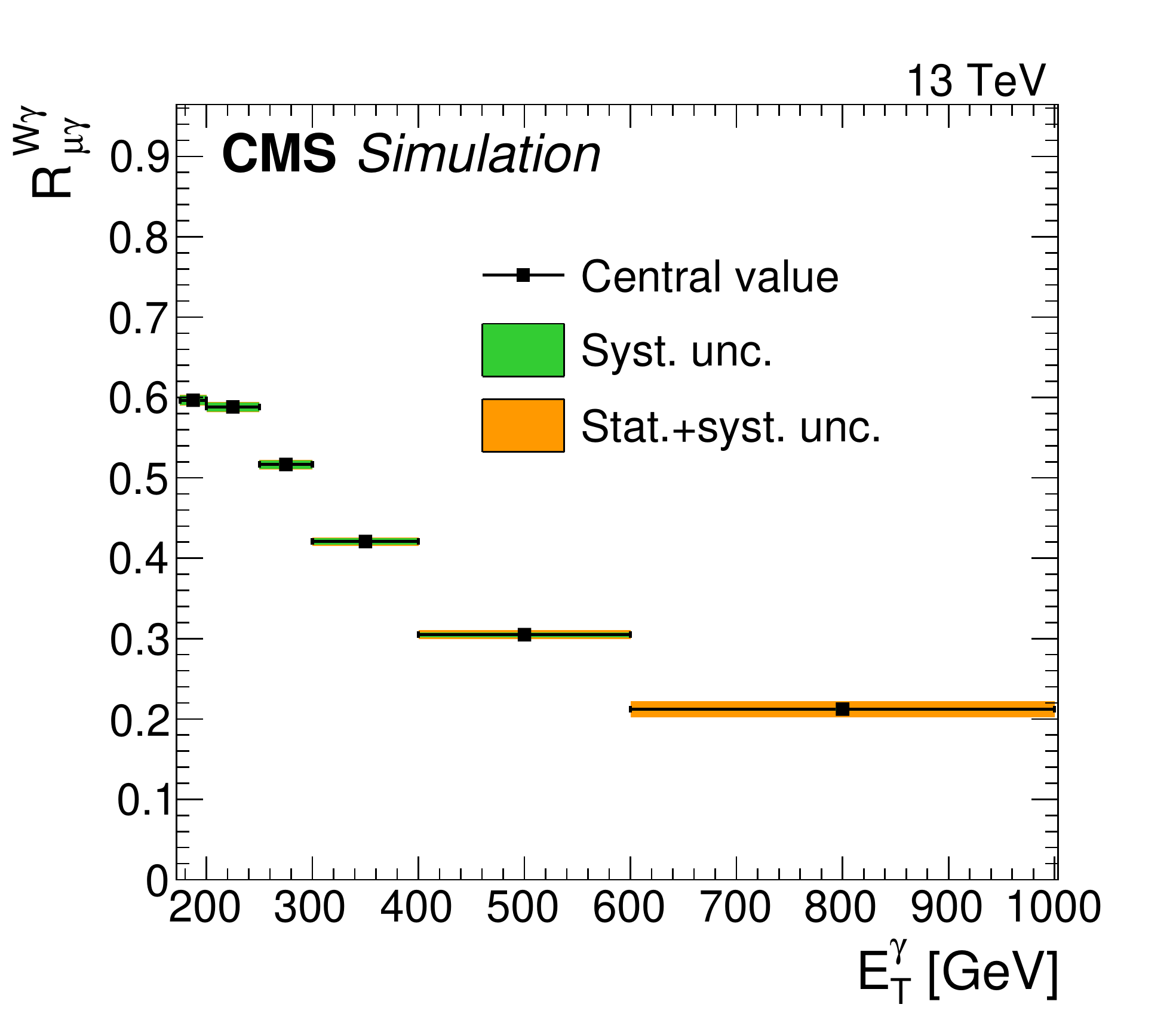}
    \caption{
      Transfer factors \RWe\ (left) and \RWm\ (right). The uncertainty bands in green (inner) and orange (outer) show the systematic uncertainty, and the combination of systematic and statistical uncertainty arising from limited MC sample size, respectively. The systematic uncertainties considered are the uncertainties in the data-to-simulation correction factors $\rho$ for the lepton identification efficiencies.
    }
    \label{fig:tf_w}
\end{figure}

Similar considerations apply to events arising from \wlng\ processes. The charged lepton from these processes
may either pass our identification criteria or fail, and in the ratio of these two classes of events, denoted
\RWl, the only uncertainties that remain non-negligible are those associated with
the lepton identification efficiency and the MC statistical uncertainty.
Figure~\ref{fig:tf_w} shows the transfer
factor \RWe\ (\RWm) between the single-electron (single-muon)
control region and the combined signal regions, for which the numerator is the estimated \wlng\ yield in the
combined signal regions, and the denominator is the estimated \wlng\ yield in the relevant control region.

Finally, an additional transfer factor $\fZW = \NZg / \NWg$ is defined to connect the \zinvg\ and
\wlng\ background yields in the signal regions, to benefit further from the larger statistical power
that the single-lepton control samples provides. The quantity $\NWg$ is the number of \wlng\ events
in the combined signal regions. When calculating the ratio \fZW, all experimental uncertainties
associated with the data-to-simulation correction factors $\rho$ cancel since both processes result in
very similar event configurations. The main uncertainties in \fZW\ are those from higher-order
theoretical corrections. The relative magnitudes of the different theoretical uncertainties are
shown in Fig.~\ref{fig:tf_syst} in Appendix~\ref{sec:app}. Figure~\ref{fig:tf_wz} shows the
transfer factor \fZW\ between the \zinvg\ and \wlng\ processes in the combined signal region.
For every transfer factor described above, both the numerator and the denominator are estimated in MC.

For increasing \ETg, the \PZ\ boson in a \zllg\ event tends to
emerge with lower rapidity, and hence so do its decay products. As a consequence,
the charged leptons are more likely to fall within the inner tracker acceptance, which increases the dilepton control region selection
efficiency of these events. In contrast, the signal region selection efficiency of \zinvg\ events is unaffected
by the rapidity of the final state neutrinos, as long as the observed \ptmiss\ has
the appropriate magnitude and azimuthal direction. This causes the distinctive drop in the ratio \RZll\ with increasing \ETg.
Similar arguments explain the drop in \RWl\ as well as the rise in \fZW. The ratio \fZW\ rises
(rather than falls) with increasing \ETg\ because \wlng\ events have a lower (rather than higher) signal region selection efficiency
if the charged lepton falls within the tracker acceptance.

Using \RWl\ and \fZW, the total estimated event yield \Tl\ in each single-lepton
control region in the $i$th bin of the \ETg\ distribution can be expressed as
\begin{equation}
  \Tl[,i] = \frac{\NZg[i]}{\RWl[,i]\fZW[,i]} + b_{\ell\Pgg,i},
\end{equation}
where $b_{\ell\Pgg}$ is the predicted contribution from other background sources
in the single-lepton regions, namely misidentified electrons and hadrons and other
minor SM processes.

\begin{figure}[htbp]
  \centering
    \includegraphics[width=0.49\textwidth]{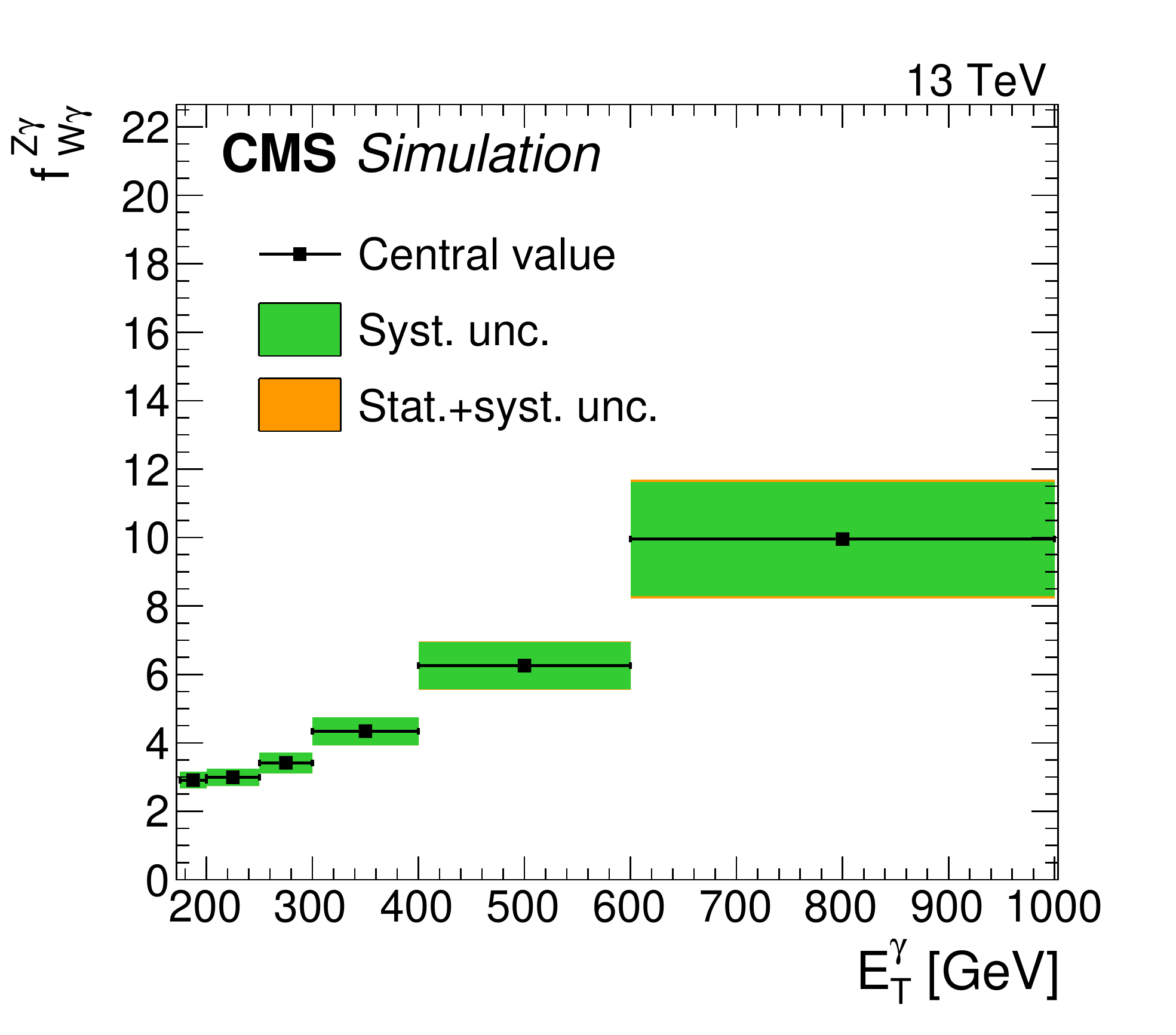}
    \caption{
      Transfer factor \fZW. The uncertainty bands in green (inner) and orange (outer) show the systematic uncertainty, and the combination of systematic and statistical uncertainty arising from limited MC sample size, respectively. The systematic uncertainties considered are the uncertainties from higher-order theoretical corrections.
    }
    \label{fig:tf_wz}
\end{figure}

\subsection{Electron misidentification background}
\label{subsec:efake}

An important background consists of $\PW\to\Pe\PGn$ events in which the electron is
misidentified as a photon. The misidentification occurs because of an inefficiency in seeding electron
tracks. A seeding efficiency of $\epsilon$ for electrons with
$\PT > 160\GeV$ is measured in data using a tag-and-probe~\cite{CMS:2011aa} technique in $\PZ\to\Pe\Pe$ events, and is
validated with MC simulation. Misidentified electron events are modeled by a proxy sample of
electron events, defined in data by requiring an ECAL cluster with a pixel seed.
The proxy events must otherwise pass the same criteria used to select signal candidate events. The number of
electron proxy events is then scaled by $R_{\epsilon} = (1-\epsilon)/\epsilon$ to yield an estimated contribution
of events from electron misidentification to our signal candidate selection. The ratio $R_{\epsilon}$ was measured to be
$0.0303 \pm 0.0022$ and uniform across the considered \ETg\ spectrum, with the dominant uncertainty in this
estimate coming from the statistical uncertainty in the measurement of $\epsilon$.

\subsection{Jet misidentification background}
\label{subsec:jfake}

Electromagnetic showers from hadronic activity can also mimic a photon signature. This process is
estimated by counting the numbers of events in two different subsets of a low-\ptmiss\ multijet data
sample. The first subset consists of events with a photon candidate that satisfies the signal selection
criteria. These events contain both genuine photons and jets that are misidentified as photons. The
second subset comprises events with a candidate photon that meets less stringent shower shape
requirements and inverted isolation criteria with respect to the signal candidates. Nearly all of
the candidate photons in these events arise from jet misidentification. The hadron misidentification
ratio is defined as the ratio between the number of misidentified events in the first subset to
the total number of events in the second subset.

The numerator is estimated by fitting the observed shower
shape distribution of the photon candidate in the first subset with a combination of simulated distributions
and distributions obtained from the observed data. For genuine
photons, the shower width distribution is formed using simulated $\Pgg + \text{jets}$ events.  For
jets misidentified as photons, the distribution is obtained from a sample selected by inverting the
charged-hadron isolation and removing the shower-shape requirement entirely.

The hadron misidentification ratio is measured to be between 0.08 and 0.12 with a few percent
relative uncertainty depending on the energy of the photon candidate. The dominant uncertainty is
systematic, and comprises the shower shape distribution fit and shower shape modelling uncertainty,
along with uncertainties associated with variations in the charged hadron isolation threshold,
low-\ptmiss\ requirement, and template bin width.

The final estimate of the
contribution of jet misidentification background to our signal candidate selection is computed by multiplying
the hadron misidentification ratio by the number of events in the high-\ptmiss\ control sample with a photon
candidate that satisfies the conditions used to select the second subset of the low-\ptmiss\ control sample.

\subsection{Beam halo and spikes background}
\label{subsec:halo}
Estimates of beam halo background and spike background are derived from fits of the
angular and timing distributions of the calorimeter clusters. Energy clusters in the ECAL due to beam
halo muons are observed to concentrate around $\abs{\sin(\phi)} \sim 0$, while all other
processes (collision-related processes and ECAL spikes) produce photon candidates that are
uniformly distributed in $\phi$~\cite{Beamhalo}, motivating the splitting of the signal region introduced in
Section~\ref{sec:event_selection}.

The splitting of the signal region can be thought of as a two-bin fit. Collision processes occupy the relative
fractions of phase space in the horizontal ($H$) and  vertical ($V$) signal regions, $C_{H} = 1/\pi$ and $C_{V} = (\pi-1)/\pi$,
respectively. The corresponding fractions for beam halo events are determined by selecting a halo-enriched sample
where the halo identification is inverted. Thus, a fit of the two signal regions provides an estimate of the overall
normalization of the beam halo background, denoted $h$. The \ETg\ dependence of the halo background is encoded in
\nhalo[,i], the unit-normalized beam halo prediction in bin $i$ of the signal region $K \in \{H,V\}$.
Using the notation introduced in Section~\ref{subsec:vgamma}, the total estimated background \TK\ in the two signal regions
are
\begin{equation*}
\begin{aligned}
  \TK[,i] & = C_{K} (\NZg[i] + \NWg[i]) + h \nhalo[,i] + C_{K} b_{K,i} \\
          & = C_{K} (1 + {\fZW[i]}^{-1}) \NZg[i] + h \nhalo[,i] + C_{K} b_{K,i},
\end{aligned}
\end{equation*}
where $b_{K,i}$ is the total contribution to bin $i$ of region $K$ from electron and hadron misidentification, ECAL spikes, and
other minor SM background processes.

The distribution of the cluster seed timing provides a cross-check on the beam halo background estimate
and an independent means to estimate the ECAL spikes contribution~\cite{Spike}. A three-component fit of the cluster
seed timing using the halo, spike, and prompt-photon templates are performed. The timing distribution of
the spike background is obtained by inverting the lower bound on the shower shape requirement in the
candidate photon selection. A total spike background of $22.9 \pm 5.8$ events is predicted, where the
dominant uncertainty is statistical.

\subsection{Other minor SM background processes}
The SM \ttg, VV\Pgg, \zllg, $\PW\to\ell\PGn$, and \gj\ processes are minor ($\sim$10\%)
background processes in the signal region. Although \zllg\ and \gj\ do not involve high-\PT invisible
particles, the former can exhibit large \met when the leptons fail to be reconstructed, and the latter
when jet energy is severely mismeasured. The estimates for all five processes are taken from
\MGvATNLO simulations at LO in QCD and can be found in Tables~\ref{tab:yield_mask_horizontal}
and~\ref{tab:yield_mask_vertical}.

\section{Results}
\label{sec:results}

\subsection{Signal extraction}
\label{sec:extraction}

The potential signal contribution is extracted from the data via simultaneous fits to the
\ETg\ distributions in the signal and control regions. Uncertainties in various
quantities are represented by nuisance parameters in the fit. Predictions for \zinvg, \wlng, and
the beam halo backgrounds are varied in the fit. Beam halo is not a
major background, but the extraction of its rate requires a fit to the observed distributions in the
signal region.

Free parameters of the fit are the yield of \zinvg\ background in each bin of the signal regions (\NZg[i])
and the overall normalization of the beam halo background ($h$). Bin-by-bin yields of \wlng\ and
\zllg\ samples in all regions are related to the yield of \zinvg\ through the MC prediction through
the transfer factors defined in Section~\ref{subsec:vgamma}. The transfer factors are allowed to shift within the
aforementioned theoretical and experimental uncertainties.

The background-only likelihood that is maximized in the fit is
\begin{equation*}
\begin{aligned}
  \mathcal{L} & = \prod_{i} \left\{ \mathcal{L_{\text{signal}}} \, \mathcal{L_{\text{single-lepton}}} \, \mathcal{L_{\text{dilepton}}} \right\} \, \mathcal{L_{\text{nuisances}}} \\
  & = \prod_{i} \left\{
    \prod_{K=H,V} \mathcal{P}\left( d_{K, i} \left| \TK[,i] (\vec{\theta} \right.) \right) \, \prod_{\ell=\Pe,\Pgm} \mathcal{P}\left( d_{\ell\Pgg, i} \left| \Tl[,i] (\vec{\theta}) \right. \right)
    \, \prod_{\ell=\Pe,\Pgm} \mathcal{P}\left( d_{\ell\ell\Pgg, i} \left| \Tll[,i] (\vec{\theta}) \right. \right)
    \right\}  \, \prod_{j} \mathcal{N}(\theta_j) \\
  & = \prod_{i} \left\{
  \begin{gathered}
    \prod\limits_{K=H,V} \mathcal{P}\biggl( d_{K, i} \biggl| \left(1 + {\fZW[,i]}^{-1}(\vec{\theta})\right) C_{K} \NZg[i] + h \nhalo[,i](\vec{\theta}) + C_{K} b_{K, i}(\vec{\theta}) \biggr. \biggr) \\
    \times \prod\limits_{\ell=\Pe,\Pgm} \mathcal{P}\biggl( d_{\ell\Pgg, i} \biggl| \frac{\NZg[i]}{\RWl[,i](\vec{\theta}) \fZW[,i] (\vec{\theta})} + b_{\ell\Pgg, i}(\vec{\theta}) \biggr. \biggr) \\
    \times \prod\limits_{\ell=\Pe,\Pgm} \mathcal{P}\biggl( d_{\ell\ell\Pgg, i} \biggl| \frac{\NZg[i]}{\RZll[,i](\vec{\theta})} + b_{\ell\ell\Pgg, i}(\vec{\theta}) \biggr. \biggr)
  \end{gathered} \right\}
  \, \prod_{j} \mathcal{N}(\theta_j),
\end{aligned}
\end{equation*}
following the notation introduced in Section~\ref{sec:bkgestimate}, and where $\mathcal{P}
(n\vert\lambda)$ is the Poisson probability of $n$ for mean $\lambda$, $\mathcal{N}$ denotes the
unit normal distribution, and $d_{X,i}$ is the observed number of events in bin i of region X.
Systematic uncertainties are treated as nuisance parameters in the fit and are represented by
$\vec{\theta}$.  Each quantity $Q_{j}$ with a nominal value $\overline{Q}_{j}$ and a standard deviation
of the systematic uncertainty $\sigma_{j}$ appears in the likelihood function as $\overline{Q}_{j}\exp
(\sigma_{j}\theta_{j})$.

The systematic uncertainties considered in this analysis, including the ones already mentioned in Section~\ref{sec:bkgestimate}, are:
\begin{itemize}
  \item Theoretical uncertainties in \vg\ differential cross sections, incorporated as uncertainties on the transfer factors (see Section~\ref{subsec:vgamma})
  \item Uncertainties in trigger efficiency and photon and lepton identification efficiencies
  \item Electron and jet misidentification rate uncertainties (see Sections~\ref{subsec:efake} and \ref{subsec:jfake})
  \item Photon and jet energy scale uncertainties (see Refs.~\cite{Khachatryan:2016hje} and \cite{Khachatryan:2016kdb})
  \item Beam halo and ECAL spike rate and distribution uncertainties (see Section~\ref{subsec:halo})
  \item Minor SM background cross section uncertainties
  \item Uncertainty in integrated luminosity (see Ref.~\cite{CMS:2017sdi})
\end{itemize}
Of the listed uncertainties, only the first two categories have a significant impact on the result of the signal extraction fit.

\subsection{Pre-fit and post-fit distributions}

\begin{table}
\centering
\topcaption{Expected event yields in each \ETg~bin for various background processes in the horizontal signal region.
         The background yields and the corresponding uncertainties are obtained after performing a combined fit to data in all the control samples, excluding data in the signal region.
         The observed event yields in the horizontal signal region are also reported.}
\label{tab:yield_mask_horizontal}
\begin{tabular}{ lcccccc }
\hline
\rule[-1.2ex]{0pt}{3.8ex}\ETg~[\GeVns{}]      &         [175,  200] &         [200,  250] &         [250,  300] &         [300,  400] &         [400,  600] &         [600, 1000] \\
\hline
$\PZ\Pgg$        & $  81.2 \pm   8.0 $ & $  88.2 \pm   8.4 $ & $  38.8 \pm   4.8 $ & $  26.8 \pm   3.7 $ & $   8.8 \pm   1.9 $ & $   1.4 \pm   0.7 $ \\
$\PW\Pgg$        & $  27.9 \pm   3.7 $ & $  29.9 \pm   3.9 $ & $  11.4 \pm   1.7 $ & $   6.3 \pm   1.2 $ & $   1.4 \pm   0.4 $ & $   0.1 \pm   0.1 $ \\
Misid. electrons & $  22.5 \pm   2.7 $ & $  25.7 \pm   2.7 $ & $  10.5 \pm   1.0 $ & $   8.2 \pm   0.7 $ & $   2.7 \pm   0.2 $ & $   0.5 \pm   0.0 $ \\
Misid. hadrons   & $   5.2 \pm   2.2 $ & $   9.3 \pm   1.8 $ & $   3.1 \pm   0.7 $ & $   1.0 \pm   0.3 $ & $   0.4 \pm   0.1 $ & $   0.0 \pm   0.0 $ \\
Other SM         & $  13.6 \pm   2.0 $ & $  19.6 \pm   1.3 $ & $  13.9 \pm   0.4 $ & $   4.2 \pm   0.2 $ & $   0.8 \pm   0.0 $ & $   0.1 \pm   0.0 $ \\
ECAL spikes      & $   4.3 \pm   1.3 $ & $   2.7 \pm   0.8 $ & $   0.5 \pm   0.1 $ & $   0.1 \pm   0.0 $ & $   0.0 \pm   0.0 $ & $   0.0 \pm   0.0 $ \\[\cmsTabSkip]
Total prediction & $ 154.6 \pm   8.3 $ & $ 175.4 \pm   8.8 $ & $  78.2 \pm   5.3 $ & $  46.6 \pm   4.0 $ & $  14.1 \pm   2.1 $ & $   2.1 \pm   0.8 $ \\[\cmsTabSkip]
Observed         & $ 150   \pm  12   $ & $ 166   \pm    13 $ & $  76.0 \pm   8.7 $ & $  44.0 \pm   6.6 $ & $  19.0 \pm   4.4 $ & $   4.0 \pm   2.0 $ \\
\hline
\end{tabular}
\end{table}

\begin{table}
\centering
\topcaption{Expected event yields in each \ETg~bin for various background processes in the vertical signal region.
         The background yields and the corresponding uncertainties are obtained after performing a combined fit to data in all the control samples, excluding data in the signal regions.
         The observed event yields in the vertical signal region are also reported.}
\label{tab:yield_mask_vertical}
\begin{tabular}{ lcccccc }
\hline
\rule[-1.2ex]{0pt}{3.8ex}\ETg~[\GeVns{}]      &         [175,  200] &         [200,  250] &         [250,  300] &         [300,  400] &         [400,  600] &         [600, 1000] \\
\hline
$\PZ\Pgg$        & $ 172   \pm    17 $ & $ 190   \pm  18   $ & $  83   \pm  10   $ & $  58.6 \pm   7.9 $ & $  18.0 \pm   3.9 $ & $   3.1 \pm   1.6 $ \\
$\PW\Pgg$        & $  59.9 \pm   7.8 $ & $  63.6 \pm   7.8 $ & $  24.6 \pm   3.5 $ & $  13.4 \pm   2.4 $ & $   3.0 \pm   0.8 $ & $   0.3 \pm   0.2 $ \\
Misid. electrons & $  48.4 \pm   5.6 $ & $  56.2 \pm   5.1 $ & $  23.4 \pm   1.8 $ & $  15.7 \pm   1.4 $ & $   5.6 \pm   0.4 $ & $   1.2 \pm   0.1 $ \\
Misid. hadrons   & $  15.1 \pm   4.4 $ & $  14.5 \pm   3.1 $ & $   4.2 \pm   0.8 $ & $   2.3 \pm   0.8 $ & $   0.5 \pm   0.1 $ & $   0.1 \pm   0.1 $ \\
Other SM         & $  33.8 \pm   4.1 $ & $  36.6 \pm   2.7 $ & $  13.6 \pm   0.5 $ & $  17.1 \pm   0.6 $ & $   2.4 \pm   0.1 $ & $   0.8 \pm   0.0 $ \\
ECAL spikes      & $   9.3 \pm   2.8 $ & $   5.7 \pm   1.7 $ & $   0.9 \pm   0.3 $ & $   0.3 \pm   0.1 $ & $   0.0 \pm   0.0 $ & $   0.0 \pm   0.0 $ \\[\cmsTabSkip]
Total prediction & $ 339   \pm  18  $ & $ 366   \pm  19   $ & $ 150   \pm  11   $ & $ 107.5 \pm   8.7 $ & $  29.6 \pm   4.3 $ & $   5.4 \pm   1.7 $ \\[\cmsTabSkip]
Observed         & $ 301   \pm  17   $ & $ 342   \pm  19   $ & $ 161 \pm  13   $ & $ 107   \pm  10   $ & $  41.0 \pm   6.4 $ & $  12.0 \pm   3.5 $ \\
\hline
\end{tabular}
\end{table}

\begin{figure}[htbp]
  \centering
    \includegraphics[width=0.45\textwidth]{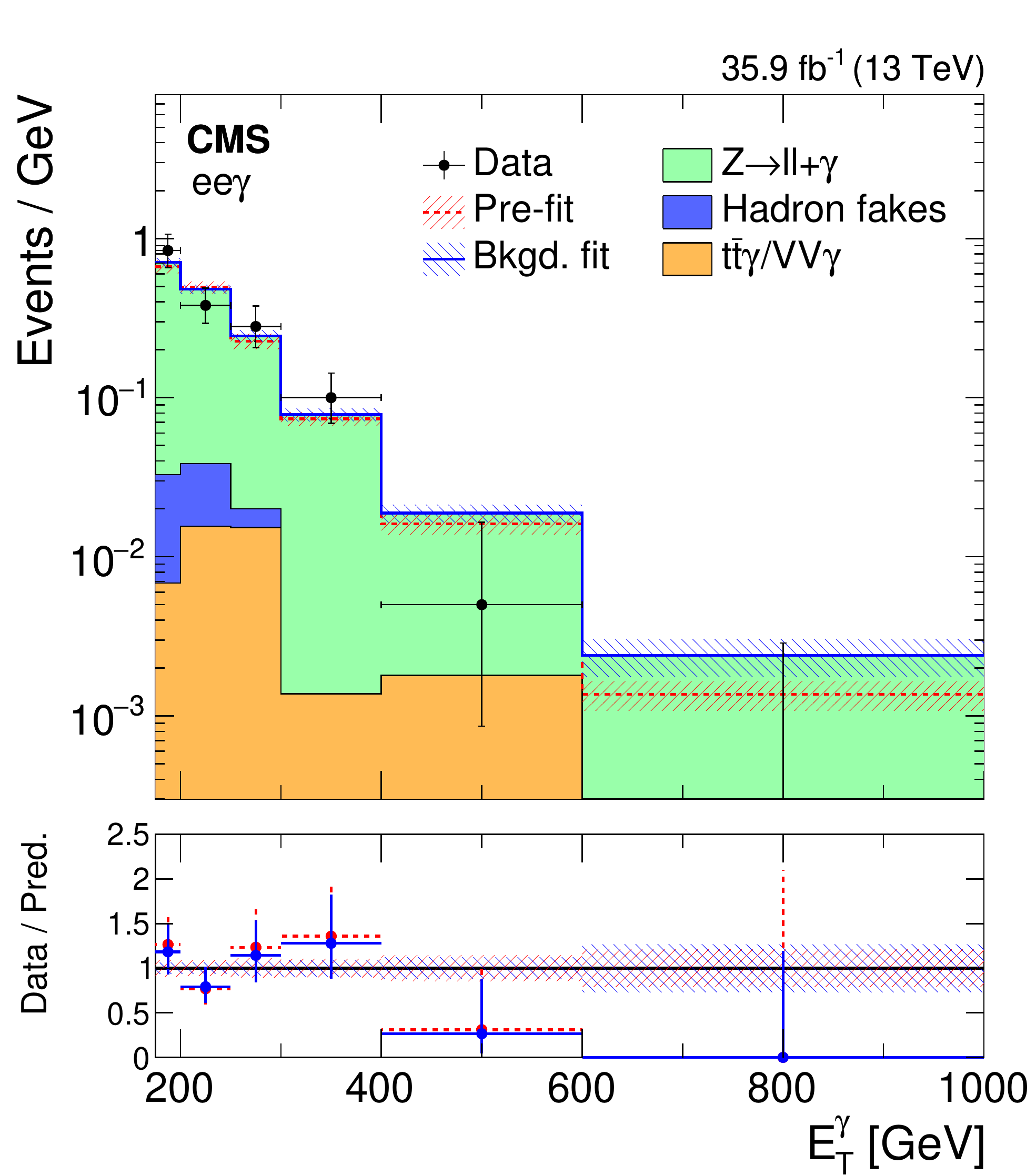}
    \includegraphics[width=0.45\textwidth]{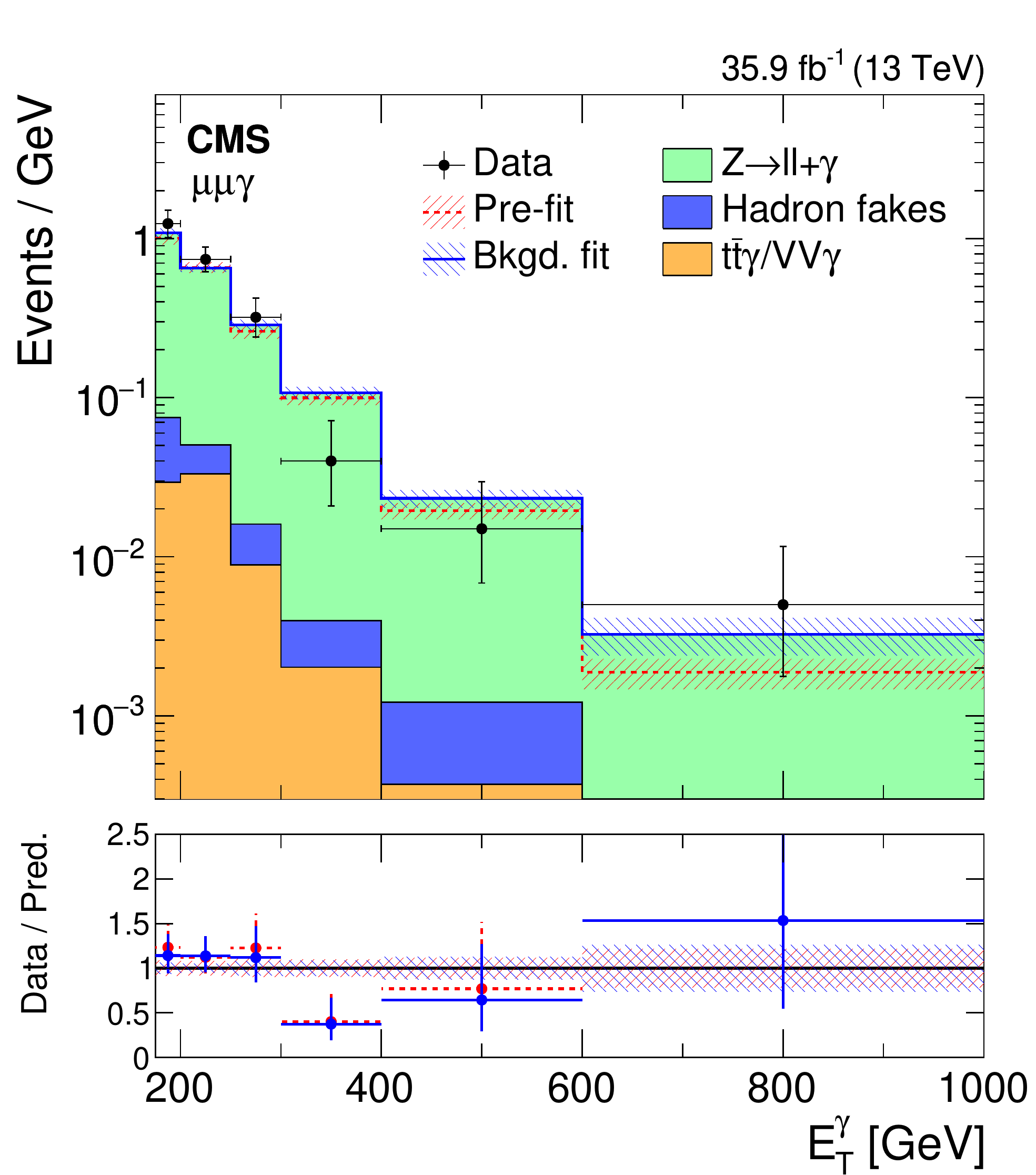}
    \includegraphics[width=0.45\textwidth]{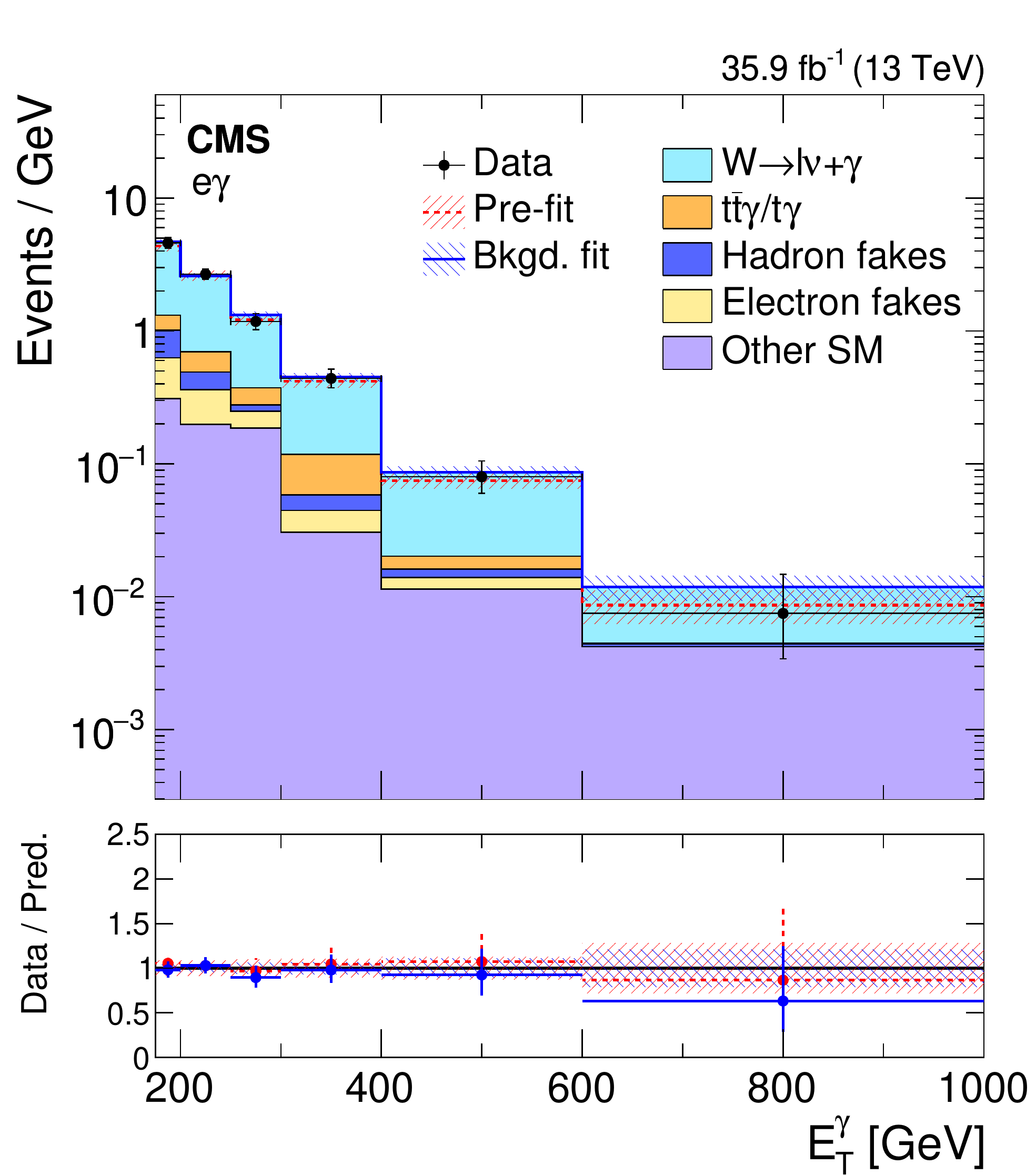}
    \includegraphics[width=0.45\textwidth]{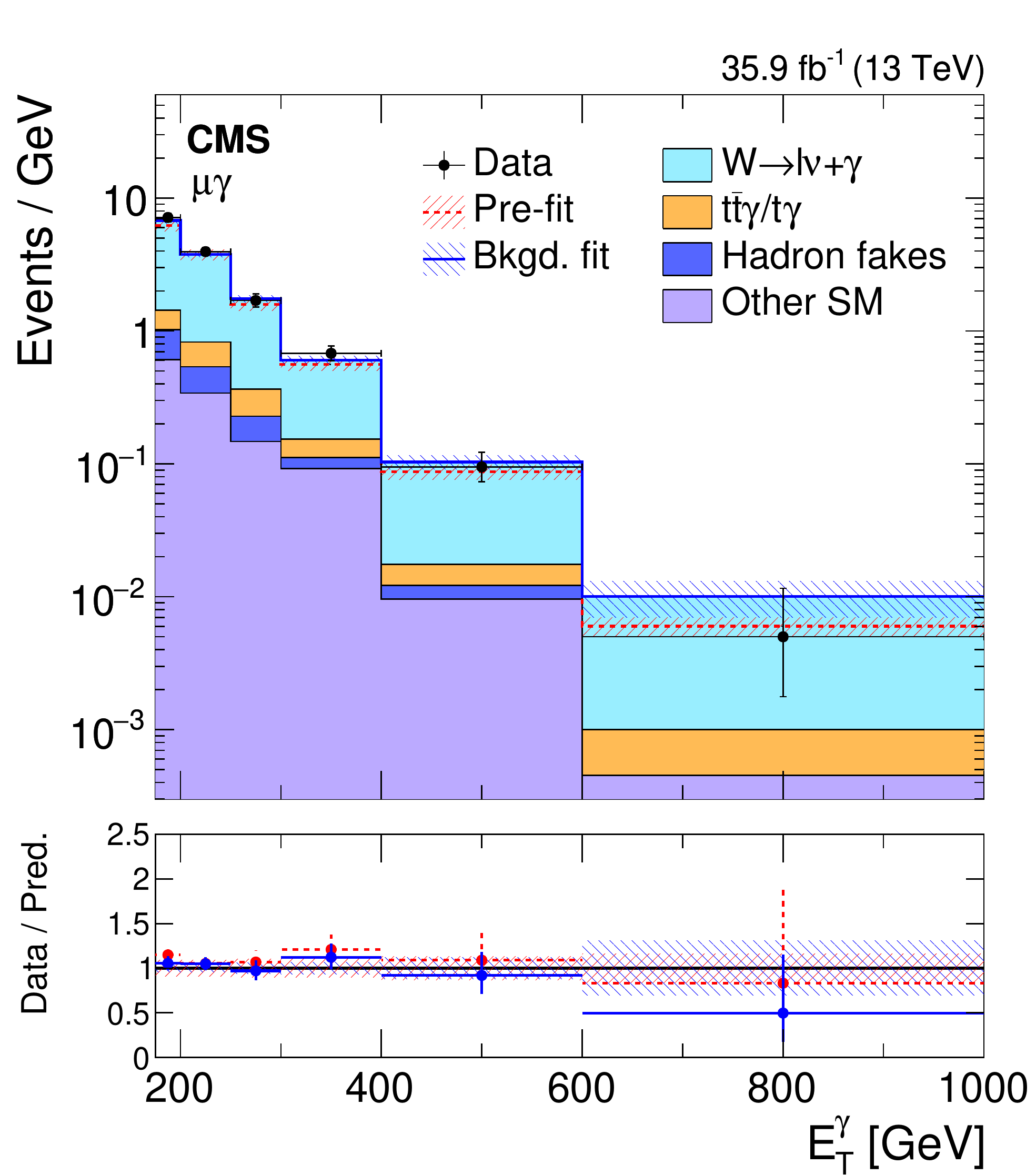}
    \caption{
      Comparison between data and MC simulation in the four control regions: \Pe\Pe\Pgg\ (upper left), \Pgm\Pgm\Pgg\ (upper right), \Pe\Pgg\ (lower left),
      \Pgm\Pgg\ (lower right) before and after performing the simultaneous fit across all the control samples and signal region, and assuming absence of any signal.
      The last bin of the distribution includes all events with $\ETg > 1000\GeV$. The ratios of data with the pre-fit background prediction (red dashed)
      and post-fit background prediction (blue solid) are shown in the lower panels. The bands in the lower panels show the post-fit uncertainty after combining all the systematic uncertainties.
}
    \label{fig:postfit1}
\end{figure}

\begin{figure}[htbp]
  \centering
    \includegraphics[width=0.45\textwidth]{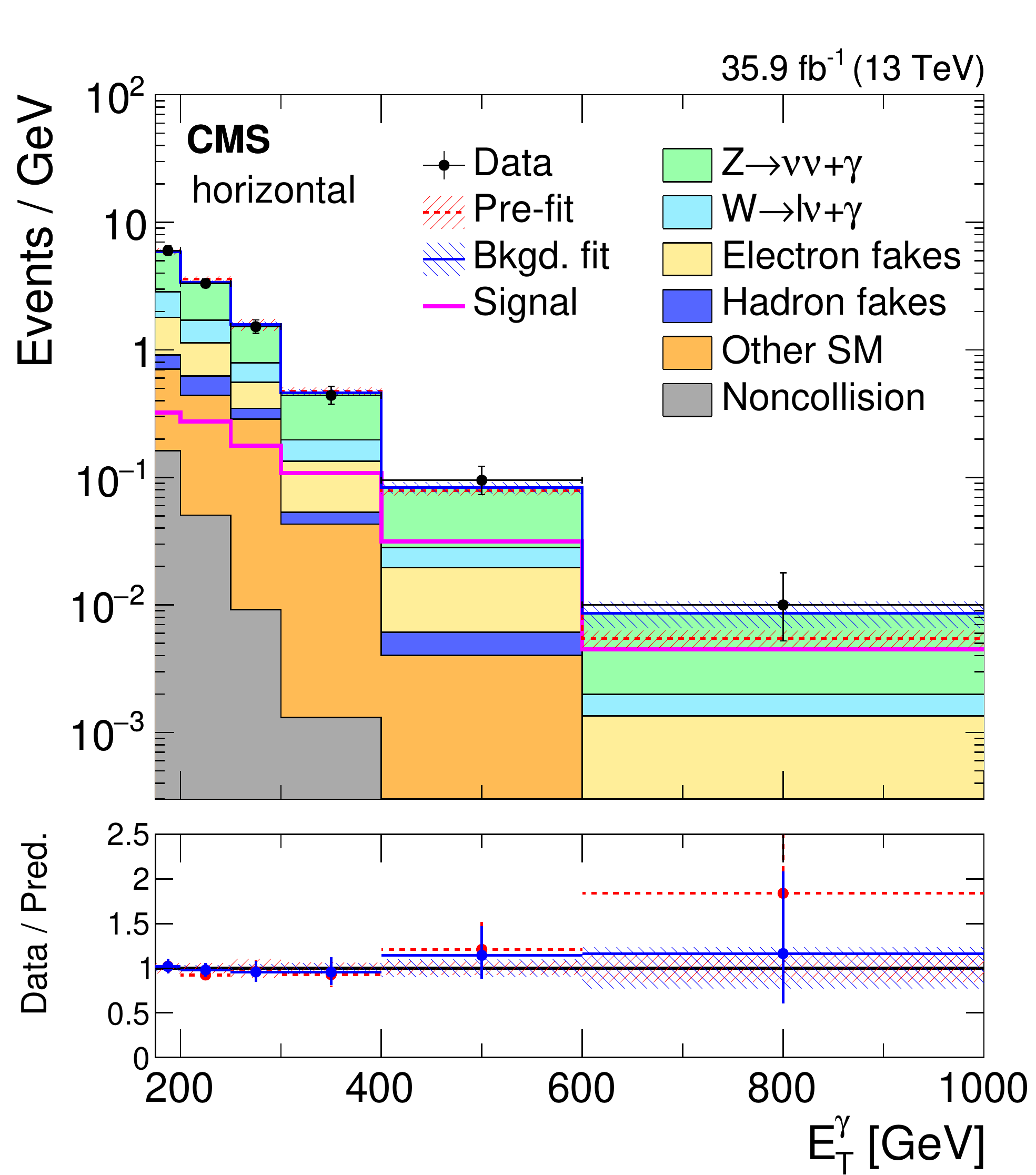}
    \includegraphics[width=0.45\textwidth]{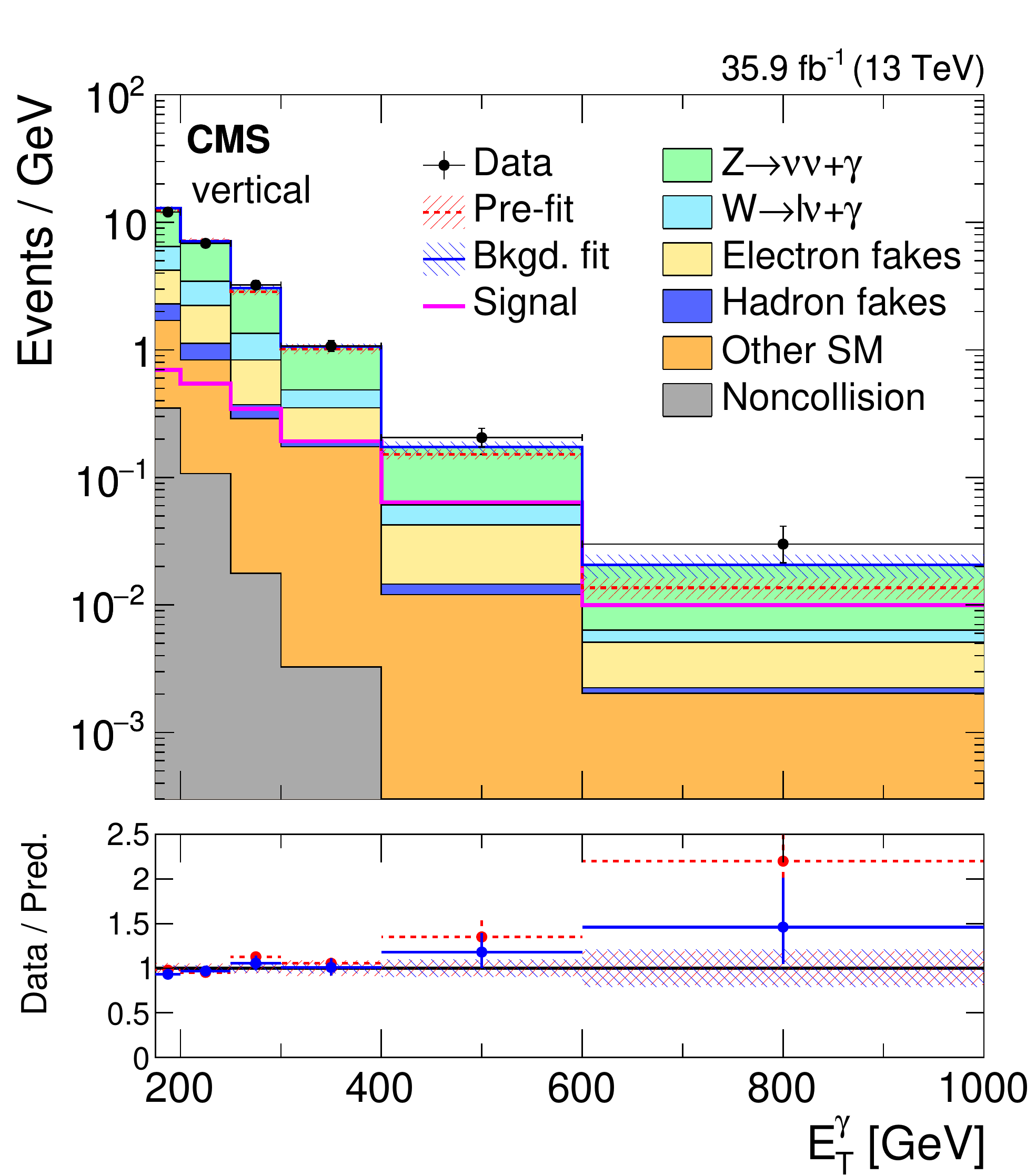}
    \caption{
      Observed \ETg\ distributions in the horizontal (left) and vertical (right) signal regions compared with the post-fit background expectations for various SM processes.
      The last bin of the distribution includes all events with $\ETg > 1000\GeV$. The expected background distributions are evaluated after performing a combined fit to
      the data in all the control samples and the signal region. The ratios of data with the pre-fit background prediction (red dashed) and post-fit
      background prediction (blue solid) are shown in the lower panels. The bands in the lower panels show the post-fit uncertainty after combining all the systematic uncertainties. The expected signal
      distribution from a 1\TeV vector mediator decaying to 1\GeV DM particles is overlaid.
    }
    \label{fig:postmonoph}
\end{figure}

 Figure~\ref{fig:postfit1} shows the observed \ETg\ distributions
in the four control regions compared with the results from simulations before and after performing the simultaneous fit across all the control samples and signal region, and assuming absence of any signal.
Figure~\ref{fig:postmonoph} shows the observed \ETg\ distributions in the horizontal and vertical
signal regions compared with the results from simulations before and after performing a combined fit to the data in all the control samples and the signal region. The observed distributions are in agreement with the prediction
from SM and noncollision backgrounds. In particular, the fit estimates the beam halo background to be zero in both regions. The dominant systematic uncertainties in the signal model include
those on the integrated luminosity, jet and $\gamma$ energy scales, \ptmiss resolution,
and data-to-simulation scale factors discussed in Section 5.

The expected yields in each bin of \ETg~for all backgrounds in the horizontal and vertical signal regions after performing
a combined fit to data in all the control samples, excluding data in the signal regions, are given in
Tables~\ref{tab:yield_mask_horizontal} and~\ref{tab:yield_mask_vertical}, respectively.
The covariances between the predicted background yields across all the \ETg~bins in the two signal regions
are shown in Fig.~\ref{fig:correlation_matrix} in Appendix~\ref{sec:app}.
The expected yields together with the covariances can be used with the simplified likelihood
approach detailed in Ref.~\cite{CMS-NOTE-2017-001} to reinterpret the results for models not studied in this paper.

\subsection{Limits}

No significant excess of events beyond the SM expectation is observed. Upper
limits are determined for the production cross section of three new-physics processes mentioned in
Section~\ref{sec:introduction}. For each model, a 95\% confidence level (\CL) upper limit is obtained utilizing
the asymptotic \CLs\ criterion~\cite{Junk:1999kv,Read:2002hq,Cowan:2010js}, using a test statistic based on the
negative logarithm of the likelihood in Section~\ref{sec:extraction}.

\begin{figure}[htbp]
  \centering
   \includegraphics[width=0.48\linewidth]{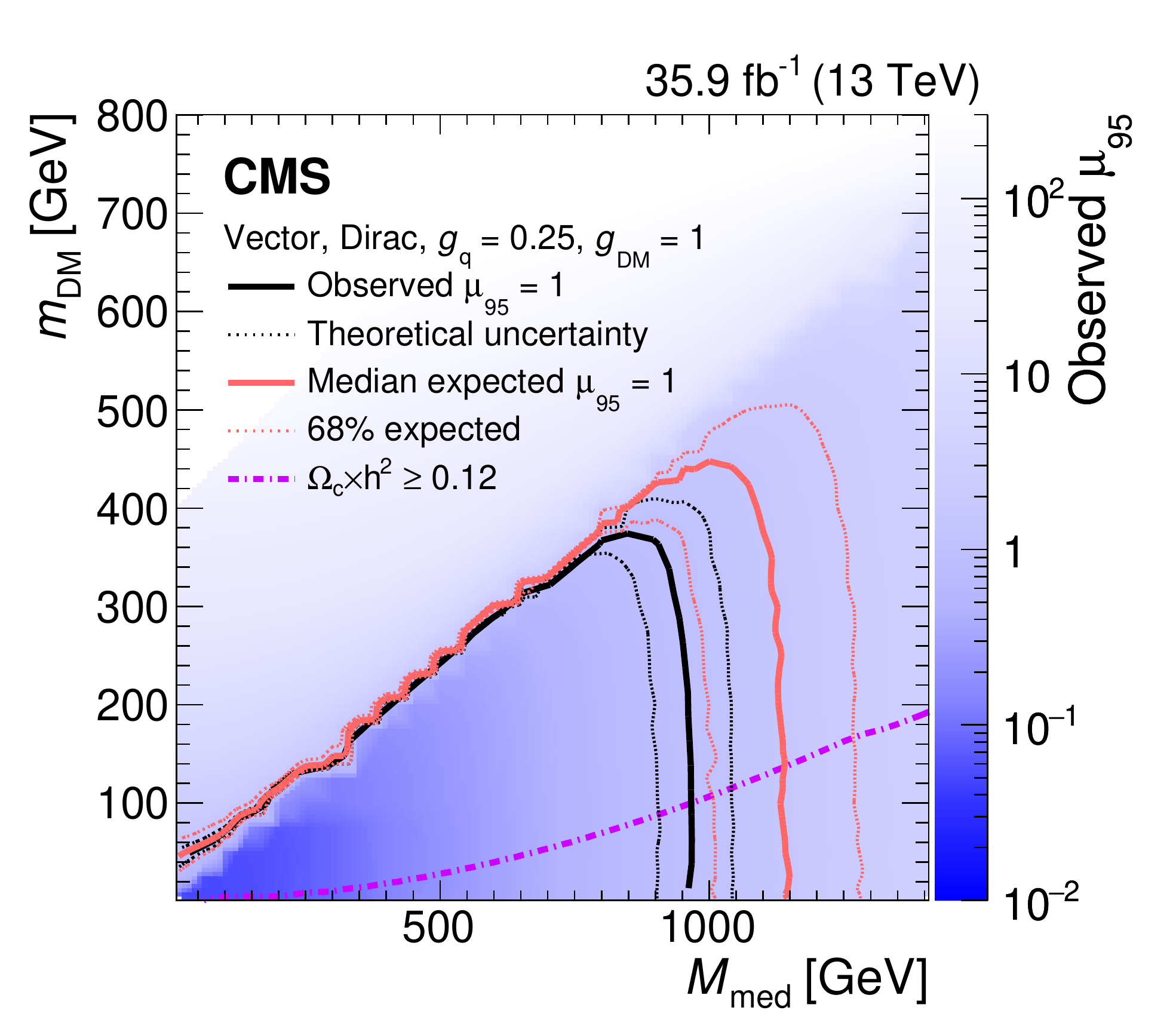}
    \includegraphics[width=0.48\linewidth]{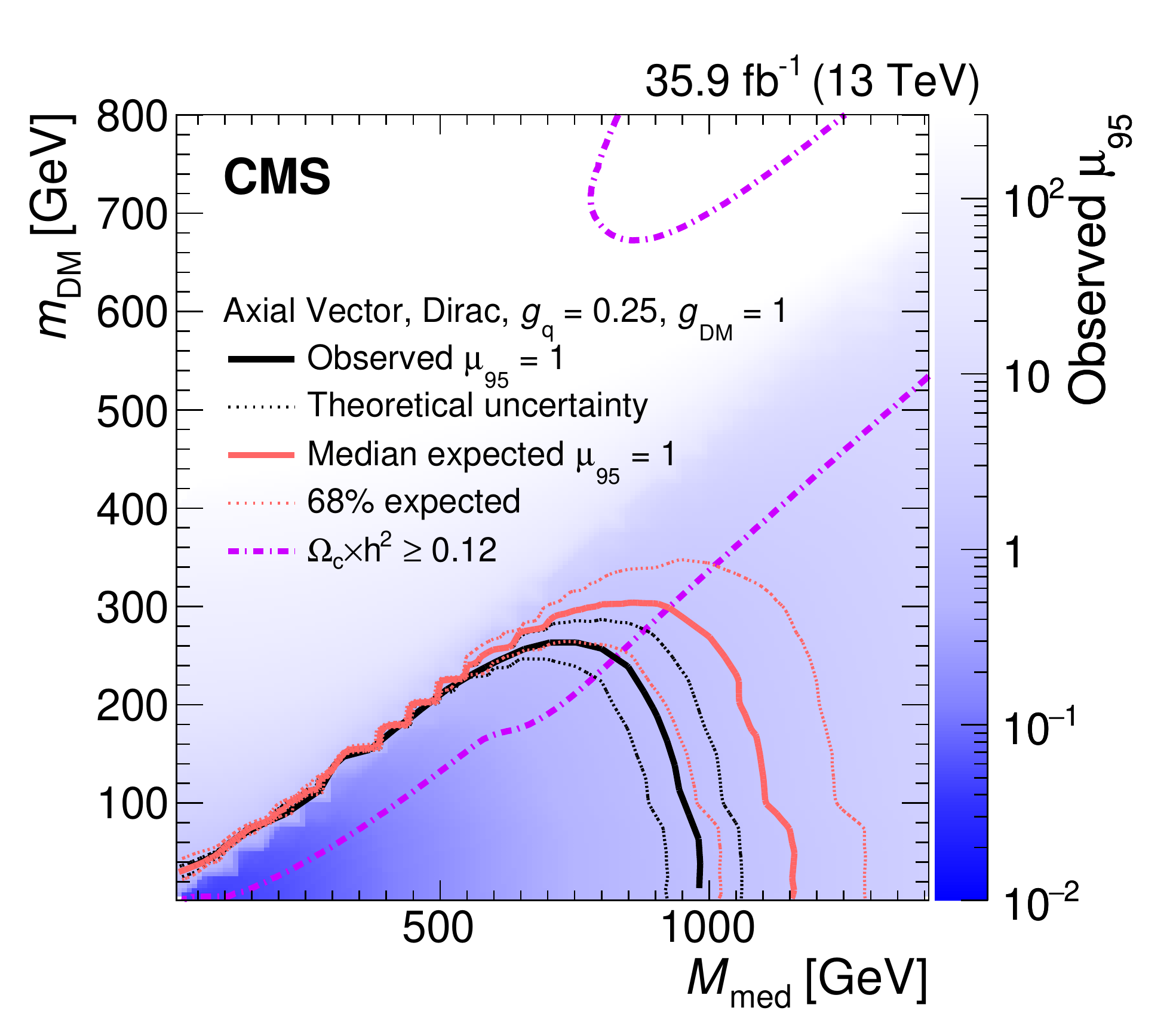}
    \caption{
      The ratio of 95\% \CL\ upper cross section limits to the theoretical cross section ($\mu_{95}$), for DM simplified models with vector (left) and axial-vector (right) mediators, assuming $\gq=0.25$ and $\gDM=1$.
      Expected $\mu_{95} = 1$ contours are overlaid in red. The region under the observed contour is excluded. For DM simplified model parameters in the region
      below the lower violet dot--dash contour, and also above the corresponding upper contour in the right hand plot, cosmological DM abundance exceeds the density observed by the Planck satellite experiment.
    }
    \label{fig:2d}
\end{figure}

The simplified DM models parameters  proposed by the ATLAS--CMS Dark Matter Forum~\cite{dmforum} are designed to
facilitate the comparison and translation of various DM search results.
Figure~\ref{fig:2d} shows the 95\% \CL\ upper cross section limits with respect to the corresponding
theoretical cross section ($\mu_{95}= \sigma_{95\%}/\sigma_{\text{theory}}$) for the  vector and
axial-vector mediator scenarios, in the \mmed--\mdm\ plane. The solid black (dashed red) curves are the
observed (expected) contours of $\mu_{95} = 1$. The $\sigma_{\text{theory}}$ hypothesis
is excluded at 95\% \CL\ or above in the region with $\mu_{95} < 1$. The uncertainty in the expected upper limit includes the
experimental uncertainties. For the simplified DM LO models considered, mediator masses up to 950\GeV are excluded for
values of \mdm\ less than 1\GeV.

The results for vector, axial-vector, and pseudoscalar mediators are compared to constraints
from the observed cosmological relic density of DM as determined from measurements of the
cosmic microwave background by the Planck satellite experiment~\cite{Ade:2015xua}.
The expected DM abundance is estimated, separately for each model, using the thermal freeze-out
mechanism implemented in the {\sc MadDM}~\cite{Backovic:2013dpa} framework and  compared to
the observed cold DM density $\Omega_c h^2=0.12$~\cite{Ade:2015xua}, where $\Omega_c$
is the DM relic abundance and $h$ is the dimensionless Hubble constant.

\begin{figure}[htbp]
  \centering
    \includegraphics[width=0.48\linewidth]{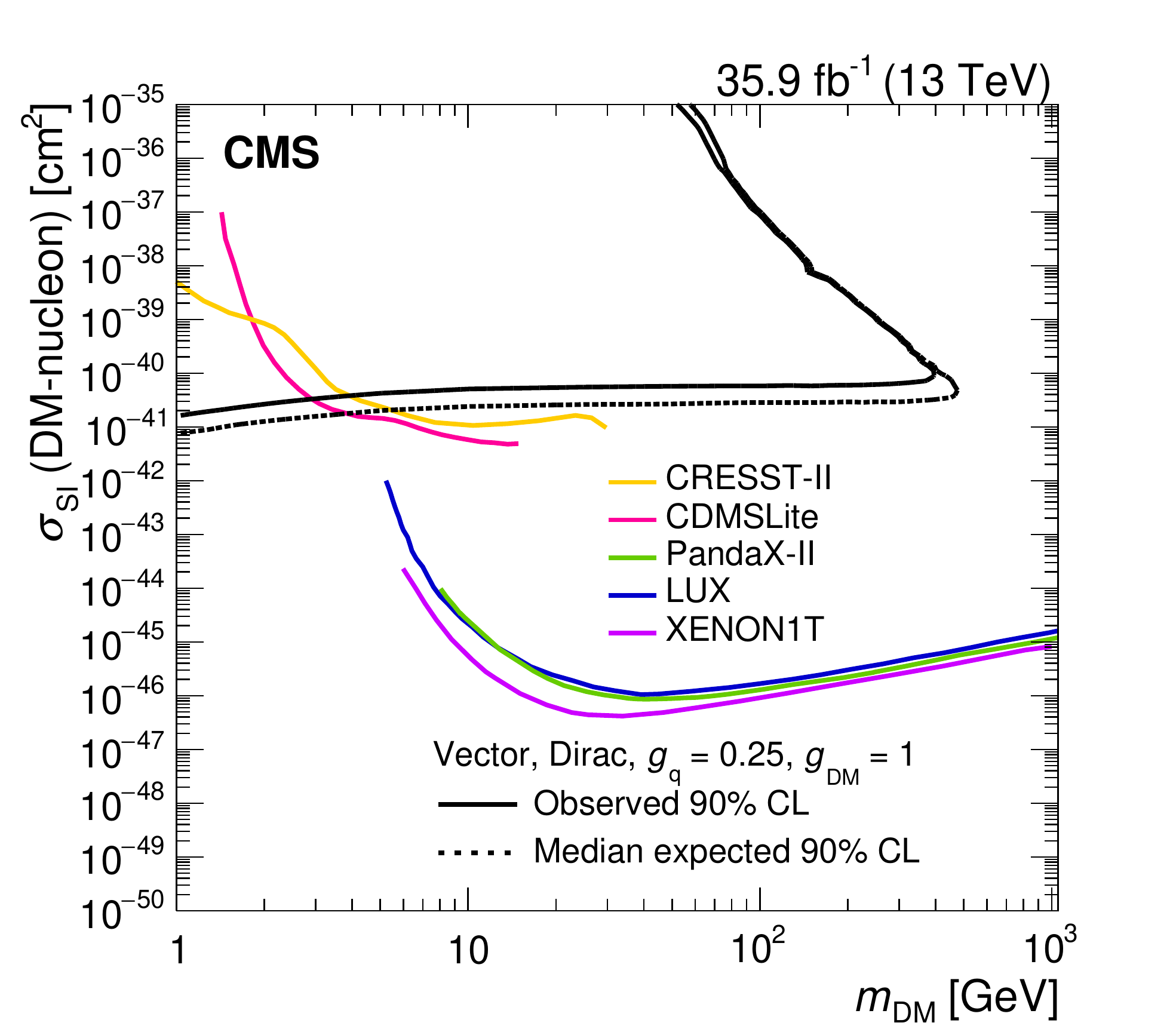}
    \includegraphics[width=0.48\linewidth]{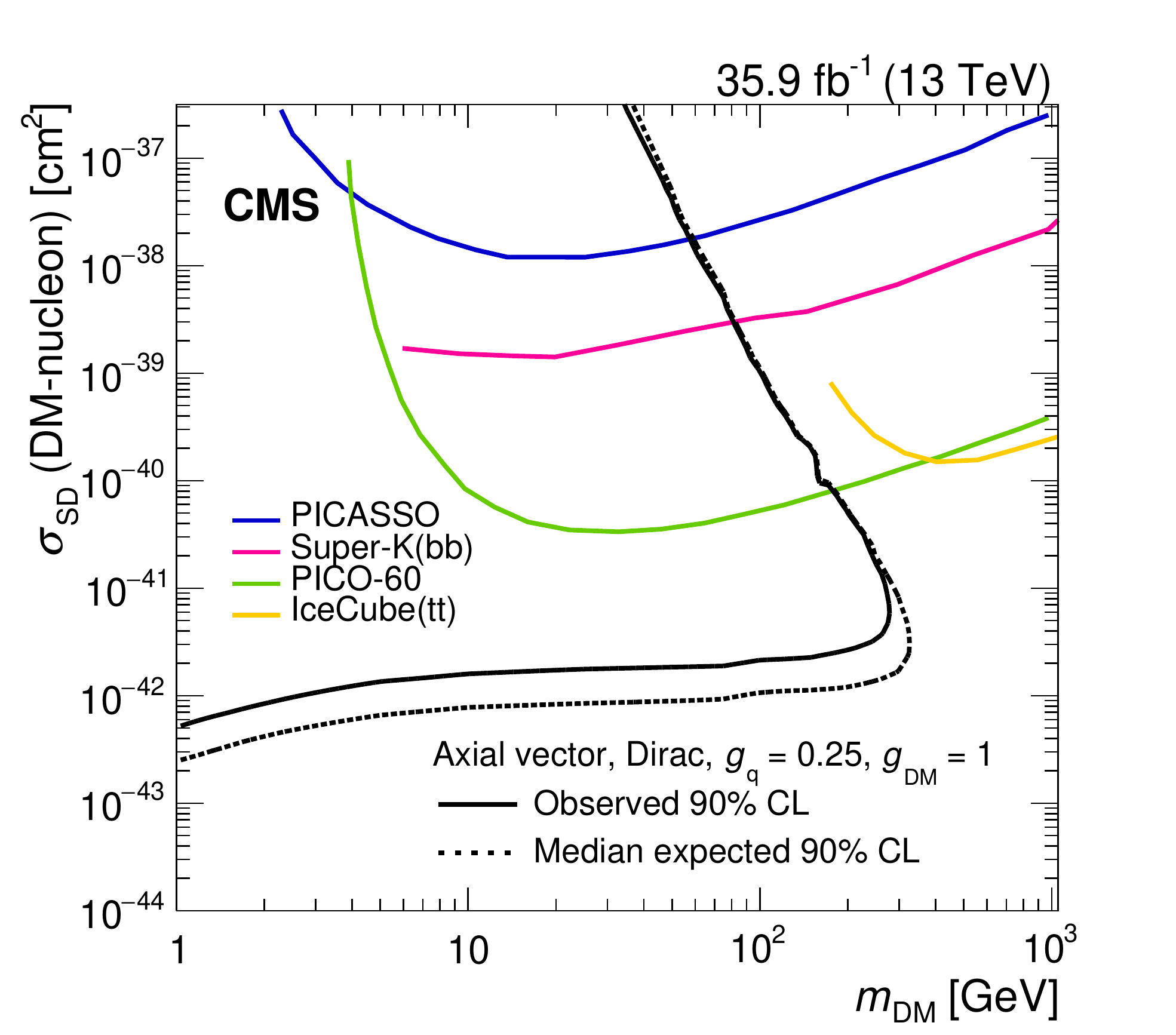}
    \caption{
      The 90\% \CL\ exclusion limits on the $\chi$--nucleon spin-independent (left)
      and spin-dependent (right) scattering cross sections involving vector and axial-vector operators, respectively, as a function of the \mdm.
      Simplified model DM parameters of $\gq=0.25$ and $\gDM=1$ are assumed.
      The region to the upper left of the contour is excluded. On the plots, the median expected 90\% \CL\ curve overlaps the observed 90\% \CL\ curve.
      Also shown are
      corresponding exclusion contours, where regions above the curves are excluded,
      from the recent results by CDMSLite~\cite{Agnese:2015nto}, LUX~\cite{Akerib:2016vxi},
      PandaX-II~\cite{Cui:2017}, XENON1T~\cite{Aprile:2018}, CRESST-II~\cite{Angloher:2015ewa}, PICO-60~\cite{Amole:2017dex},
      IceCube~\cite{Aartsen:2016exj}, PICASSO~\cite{Behnke:2016lsk} and Super-Kamiokande~\cite{Choi:2015ara} Collaborations.
    }
    \label{fig:cont}
\end{figure}

The exclusion contours in Fig.~\ref{fig:2d} are also translated into the
$\sigma_{\text{SI/SD}}$--\mdm\ plane, where $\sigma_{\text{SI/SD}}$ are the
spin-independent/spin-dependent DM--nucleon scattering cross sections as shown in Fig.~\ref{fig:cont}. The translation and presentation of the result
follows the prescription given in Ref.~\cite{Boveia:2016mrp}.
In particular, to enable a direct comparison with results
from direct detection experiments, these limits are calculated at 90\% \CL~\cite{dmforum}.
When compared to the direct detection experiments, the limits obtained from this search provide stronger constraints for DM masses less than 2\GeV (spin independent)
 and less than 200\GeV (spin dependent).

\begin{figure}[htbp]
\centering
\includegraphics[width=7.5cm,height=7.0cm]{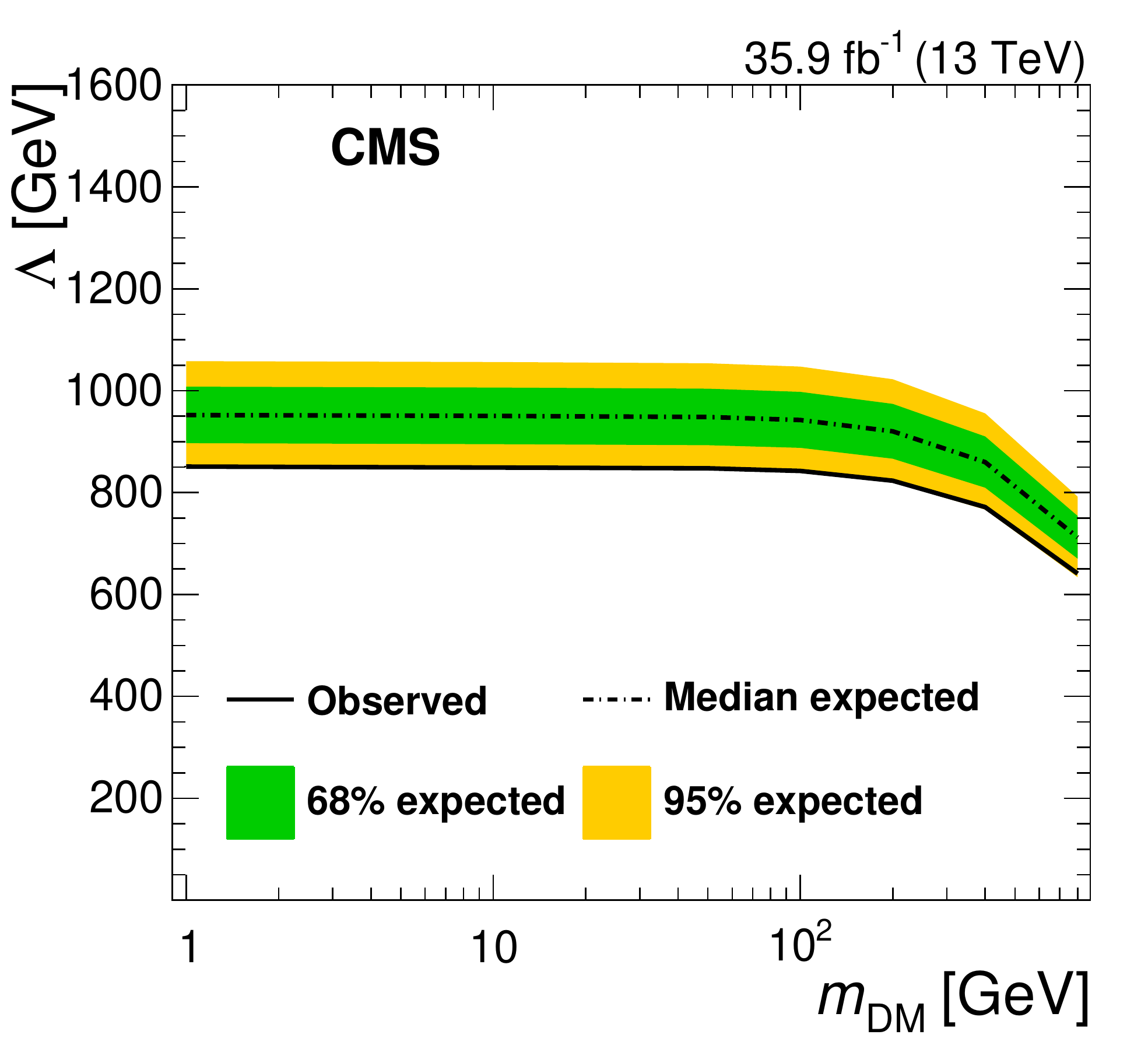}
\caption{The 95\% \CL\ observed and expected lower limits on $\Lambda$ for an effective EWK--DM contact interaction, as a function of dark matter mass \mdm.}\label{fig:DMEWKlimits}
\end{figure}

For the DM model with a contact interaction of type $\gamma\gamma\chi\overline{\chi}$, upper limits are placed on the production cross section,
which are then translated into lower limits on the suppression scale $\Lambda$ for $k_1 = k_2 = 1.0$. The 95\% \CL\ observed and expected lower limits
on $\Lambda$ as a function of dark matter mass \mdm\ are shown in Fig.~\ref{fig:DMEWKlimits}.
For \mdm\ between 1 and 100\GeV, we exclude $\Lambda$ values up to 850\,(950)\GeV, observed (expected) at 95\% \CL.

\begin{figure}[htbp]
  \centering
    \includegraphics[width=0.5\textwidth]{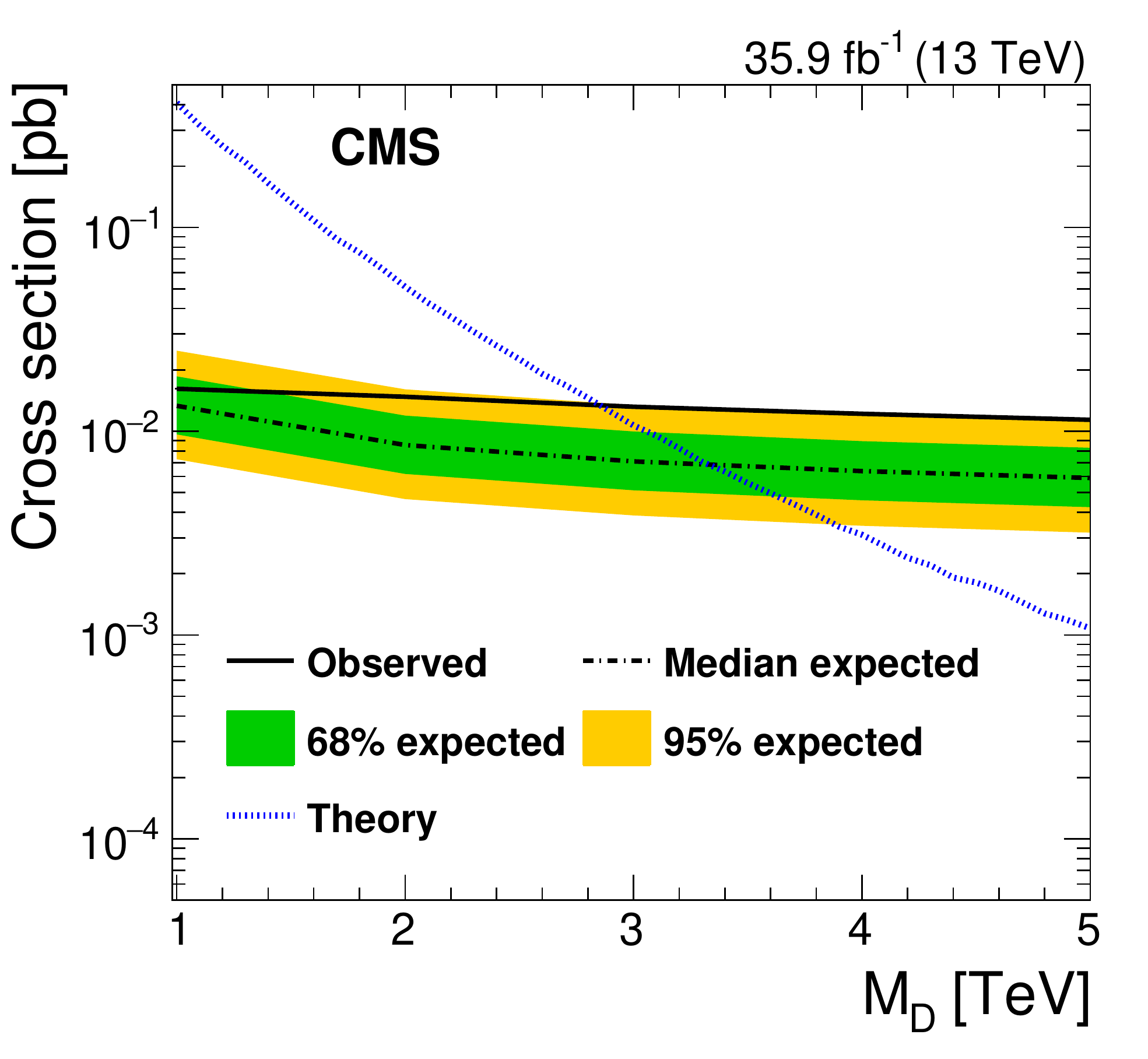}
    \caption{The 95\% \CL\ upper limits on the ADD graviton production cross section as a function of \mD, for $n=3$ extra dimensions.}
    \label{fig:ADDLimits}
\end{figure}

\begin{figure}[htbp]
  \centering
    \includegraphics[width=0.5\textwidth]{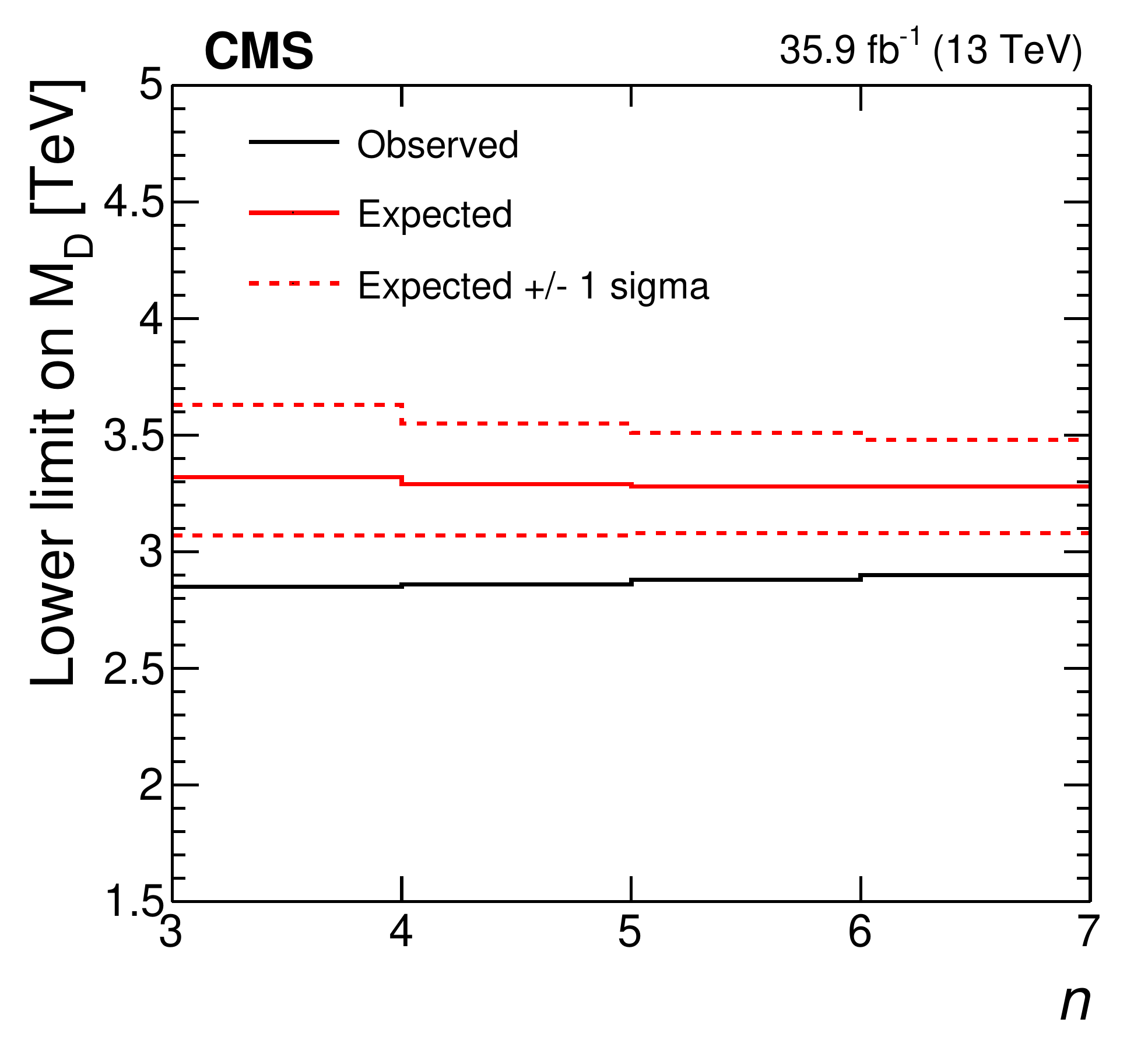}
    \caption{Lower limit on \mD\ as a function of $n$, the number of ADD extra dimensions.}
    \label{fig:MDLimits}
\end{figure}

Figure~\ref{fig:ADDLimits} shows the upper limit and the theoretically calculated ADD graviton
production cross section for $n=3$ extra dimensions, as a function of \mD. Lower limits on \mD\
for various values of $n$ extra dimensions are summarized in Table~\ref{tab:MDLimits}, and in Fig.~\ref{fig:MDLimits}.
Values of \mD\ up to 2.90\TeV for $n=6$ are excluded by the current analysis.

\begin{table}[htbp]
  \centering
    \topcaption{The 95\% \CL\ observed and expected lower limits on \mD\ as a function of $n$, the number of ADD extra dimensions.}
    \label{tab:MDLimits}
    \begin{tabular}{lcc}
      \hline
      $n$ & Obs. limit [\TeVns{}] & Exp. limit [\TeVns{}] \\
      \hline
      3 & 2.85 & 3.32 \\
      4 & 2.86 & 3.29 \\
      5 & 2.88 & 3.28\\
      6 & 2.90 & 3.28 \\
      \hline
    \end{tabular}
\end{table}

The sensitivity of the analysis to new physics, as measured by the stringency of the expected cross
section upper limits, has improved by approximately 70\% in comparison to the previous CMS
results~\cite{Sirunyan:2017ewk}. A threefold increase in the data set size accounts for one fourth
of the improvement, with the rest of the gain resulting from by the use of the simultaneous fit to
multiple signal and control regions.

\section{Summary}
\label{sec:summary}

Proton--proton collisions producing a high transverse momentum photon and large missing transverse momentum
have been investigated to search
for new phenomena, using a data set corresponding to \currentL\fbinv
of integrated luminosity recorded at $\sqrt{s} = 13\TeV$ at the
LHC. An analysis strategy of performing a simultaneous fit to
multiple signal and control regions is employed on this final state
for the first time, enhancing the sensitivity to potential signal events.
No deviations from the standard model predictions are
observed.
For the simplified
dark matter production models considered, the observed (expected) lower limit on
the mediator mass is 950\,(1150)\GeV in both cases for 1\GeV dark matter mass.
For an effective electroweak--dark matter contact interaction, the observed (expected) lower limit on
the suppression parameter $\Lambda$ is 850\,(950)\GeV.
For the model with extra
spatial dimensions, values of the effective Planck scale \mD\ up to 2.85--2.90\TeV are excluded for between 3 and 6 extra dimensions.
These limits on $\Lambda$ and \mD\ are the most sensitive monophoton limits to date.

\begin{acknowledgments}
We congratulate our colleagues in the CERN accelerator departments for the excellent performance of the LHC and thank the technical and administrative staffs at CERN and at other CMS institutes for their contributions to the success of the CMS effort. In addition, we gratefully acknowledge the computing centers and personnel of the Worldwide LHC Computing Grid for delivering so effectively the computing infrastructure essential to our analyses. Finally, we acknowledge the enduring support for the construction and operation of the LHC and the CMS detector provided by the following funding agencies: BMBWF and FWF (Austria); FNRS and FWO (Belgium); CNPq, CAPES, FAPERJ, FAPERGS, and FAPESP (Brazil); MES (Bulgaria); CERN; CAS, MoST, and NSFC (China); COLCIENCIAS (Colombia); MSES and CSF (Croatia); RPF (Cyprus); SENESCYT (Ecuador); MoER, ERC IUT, and ERDF (Estonia); Academy of Finland, MEC, and HIP (Finland); CEA and CNRS/IN2P3 (France); BMBF, DFG, and HGF (Germany); GSRT (Greece); NKFIA (Hungary); DAE and DST (India); IPM (Iran); SFI (Ireland); INFN (Italy); MSIP and NRF (Republic of Korea); MES (Latvia); LAS (Lithuania); MOE and UM (Malaysia); BUAP, CINVESTAV, CONACYT, LNS, SEP, and UASLP-FAI (Mexico); MOS (Montenegro); MBIE (New Zealand); PAEC (Pakistan); MSHE and NSC (Poland); FCT (Portugal); JINR (Dubna); MON, RosAtom, RAS, RFBR, and NRC KI (Russia); MESTD (Serbia); SEIDI, CPAN, PCTI, and FEDER (Spain); MOSTR (Sri Lanka); Swiss Funding Agencies (Switzerland); MST (Taipei); ThEPCenter, IPST, STAR, and NSTDA (Thailand); TUBITAK and TAEK (Turkey); NASU and SFFR (Ukraine); STFC (United Kingdom); DOE and NSF (USA).

\hyphenation{Rachada-pisek} Individuals have received support from the Marie-Curie program and the European Research Council and Horizon 2020 Grant, contract No. 675440 (European Union); the Leventis Foundation; the A. P. Sloan Foundation; the Alexander von Humboldt Foundation; the Belgian Federal Science Policy Office; the Fonds pour la Formation \`a la Recherche dans l'Industrie et dans l'Agriculture (FRIA-Belgium); the Agentschap voor Innovatie door Wetenschap en Technologie (IWT-Belgium); the F.R.S.-FNRS and FWO (Belgium) under the ``Excellence of Science - EOS" - be.h project n. 30820817; the Ministry of Education, Youth and Sports (MEYS) of the Czech Republic; the Lend\"ulet (``Momentum") Program and the J\'anos Bolyai Research Scholarship of the Hungarian Academy of Sciences, the New National Excellence Program \'UNKP, the NKFIA research grants 123842, 123959, 124845, 124850 and 125105 (Hungary); the Council of Science and Industrial Research, India; the HOMING PLUS program of the Foundation for Polish Science, cofinanced from European Union, Regional Development Fund, the Mobility Plus program of the Ministry of Science and Higher Education, the National Science Center (Poland), contracts Harmonia 2014/14/M/ST2/00428, Opus 2014/13/B/ST2/02543, 2014/15/B/ST2/03998, and 2015/19/B/ST2/02861, Sonata-bis 2012/07/E/ST2/01406; the National Priorities Research Program by Qatar National Research Fund; the Programa Estatal de Fomento de la Investigaci{\'o}n Cient{\'i}fica y T{\'e}cnica de Excelencia Mar\'{\i}a de Maeztu, grant MDM-2015-0509 and the Programa Severo Ochoa del Principado de Asturias; the Thalis and Aristeia programs cofinanced by EU-ESF and the Greek NSRF; the Rachadapisek Sompot Fund for Postdoctoral Fellowship, Chulalongkorn University and the Chulalongkorn Academic into Its 2nd Century Project Advancement Project (Thailand); the Welch Foundation, contract C-1845; and the Weston Havens Foundation (USA).
\end{acknowledgments}

\bibliography{auto_generated}

\appendix

\clearpage

\section{Higher-order corrections to \texorpdfstring{\vg}{V + gamma} differential cross sections}
\label{sec:app}

\begin{figure}[htbp]
  \centering
  \includegraphics[width=0.48\linewidth]{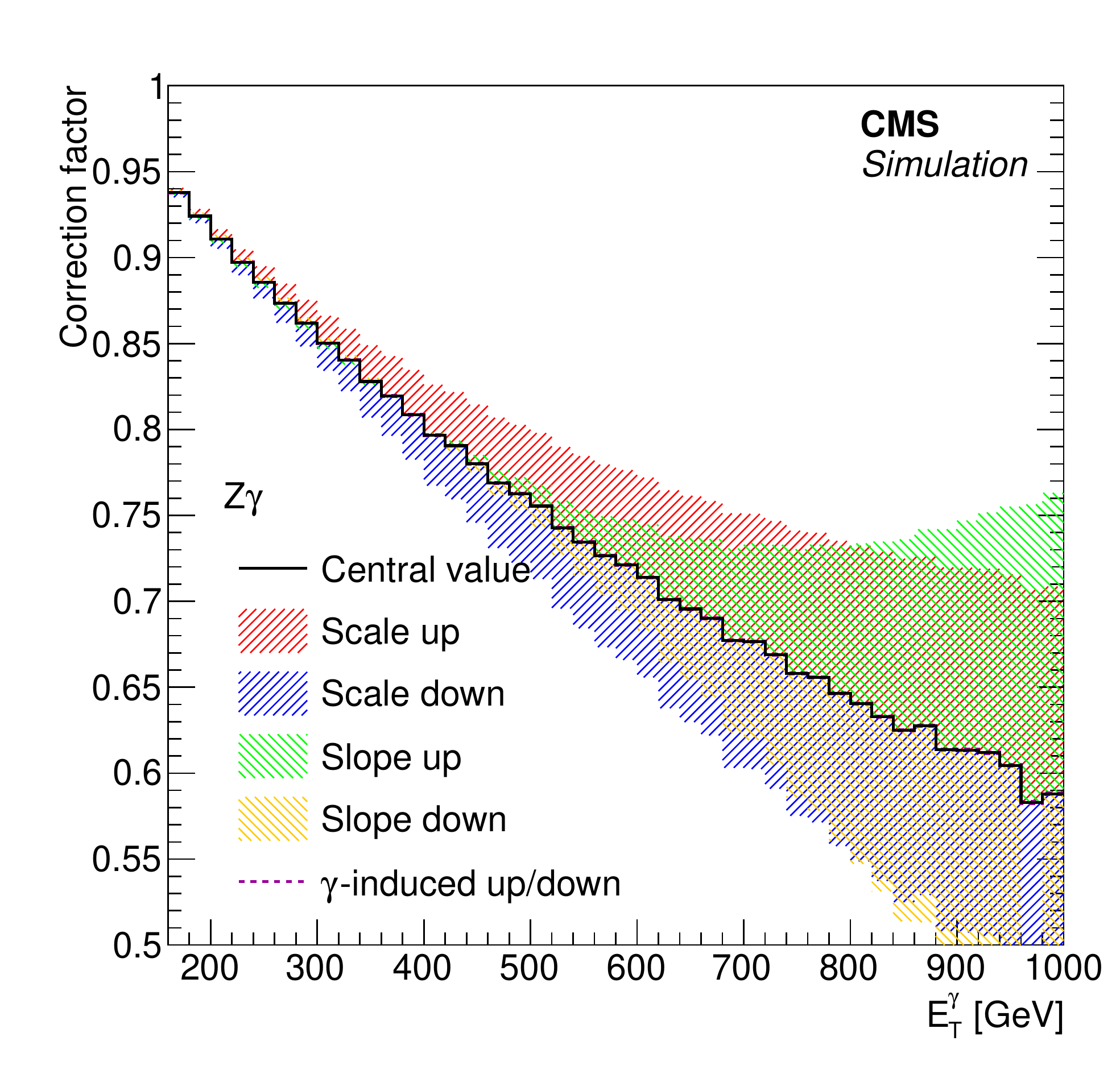} \\
  \includegraphics[width=0.48\linewidth]{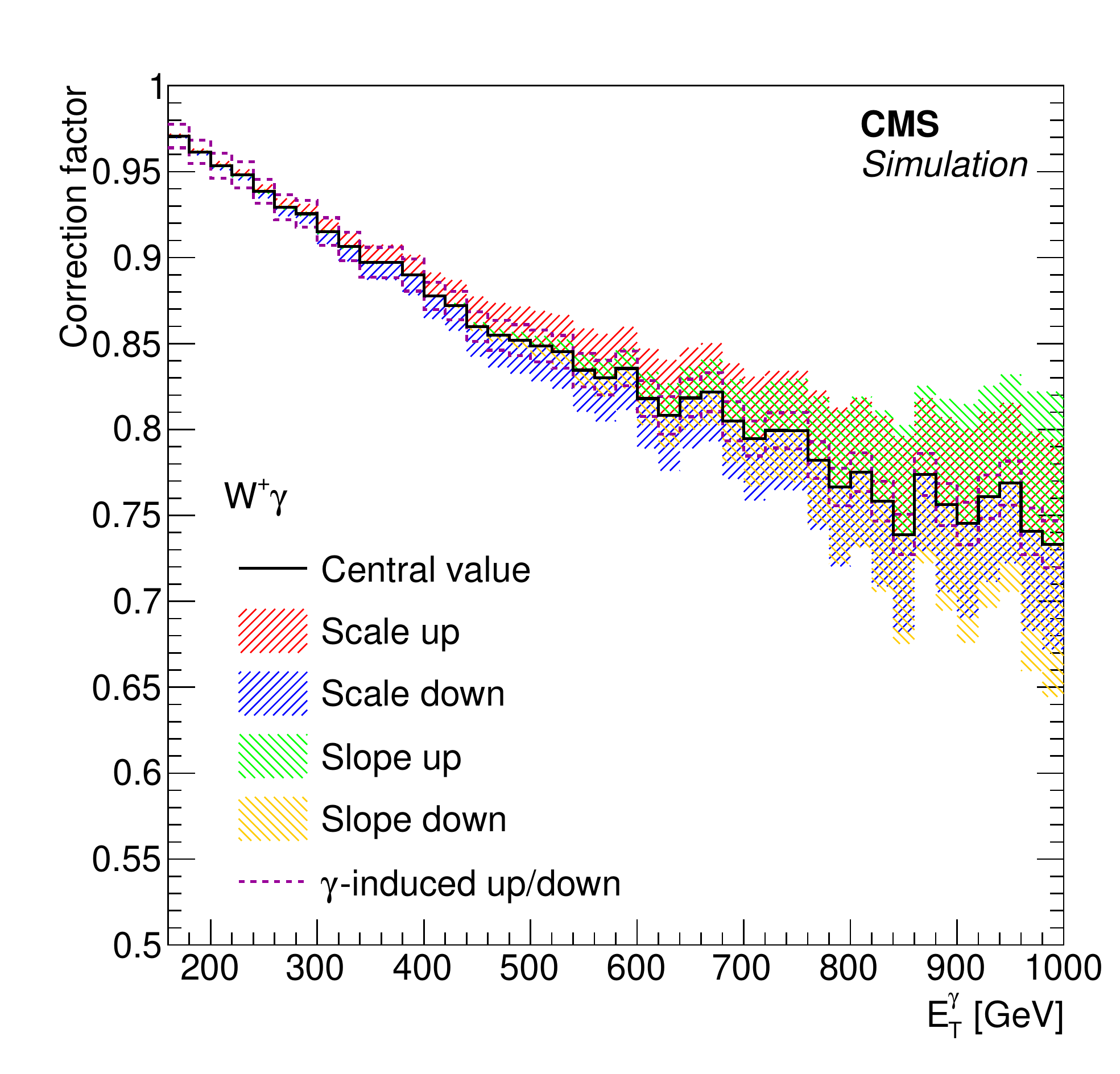}
  \includegraphics[width=0.48\linewidth]{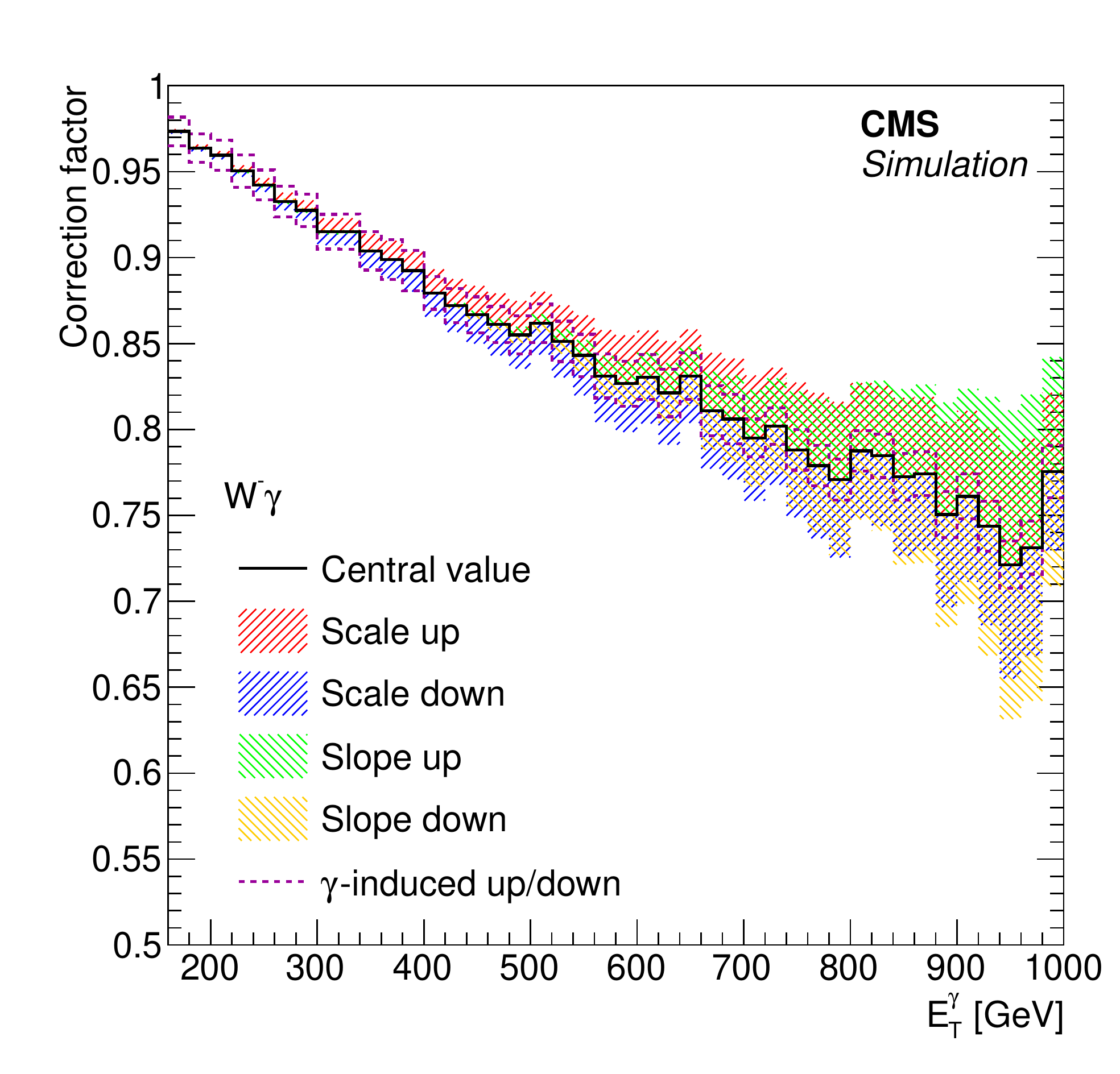}
  \caption{
    Electroweak NLO cross section corrections as a function of photon
    \PT for \zinvg\ (top), $\PW^{+}+\Pgg$ (bottom left), and $\PW^{-}+\Pgg$ (bottom right)
    processes, overlaid with uncertainty bands. See text for descriptions of the individual components of the uncertainty.
    Uncertainty due to \Pgg-induced production is negligible in \zinvg\ production.
  }
  \label{fig:ewk_correction}
\end{figure}

In order to account for higher-order electroweak corrections, we apply additional factors as a function of $\ETg$.
Of the various electroweak higher-order effects, ones that can give sizeable
($\gg\mathcal{O}(\alpha)$) corrections to the differential cross section are Sudakov suppression at high photon \PT and
potentially the addition of photon-induced scattering
processes~\cite{Denner:2014bna,Denner:2015fca}. We apply the correction factors shown in
Fig.~\ref{fig:ewk_correction}, which are combinations of Sudakov suppression factors and
photon-induced enhancements, and are provided by the authors of Ref.~\cite{Denner:2015fca} in addition to
the NNLO QCD correction.

The differential cross section after the full higher-order corrections is therefore denoted as
\begin{equation}
  \rd\sigma^{\text{NNLO QCD}+\text{NLO EW}} = \rd\sigma^{\text{LO}} k^{\text{NNLO QCD}} (1 + \kappaSudakov + \kappaPhoton),
\end{equation}
where $k^{\text{NNLO QCD}} = \rd\sigma^{\text{NNLO QCD}} / \rd\sigma^{\text{LO}}$, and the two $\kappa$
terms are the Sudakov suppression and photon-induced enhancement components of the electroweak
correction, respectively.

 We estimate the magnitude of the uncertainty in \kappaSudakov\ and \kappaPhoton\ to be $(\kappaSudakov)^2$ and
\kappaPhoton, \ie, square of the correction for Sudakov suppression and the 100\% of the correction itself for the photon-induced
enhancement. The choice of using the square of \kappaSudakov\ is motivated by the fact
that fully resummed leading-log Sudakov suppression is an exponential of \kappaSudakov.

For the Sudakov suppression, which is the dominant term in the electroweak correction, we further
consider two types of systematic variations, inspired by Ref.~\cite{Lindert:2017olm}, which provides
a prescription for electroweak correction uncertainties for $\mathrm{V}+\text{jets}$ processes. In
that paper, electroweak correction as a function of the boson \PT is varied in overall scale and in
slope. The slope variation is realized by selecting a point in the boson \PT spectrum and letting
the shift in correction cross over at the point.

\begin{figure}[htbp]
  \centering
    \includegraphics[width=0.45\textwidth]{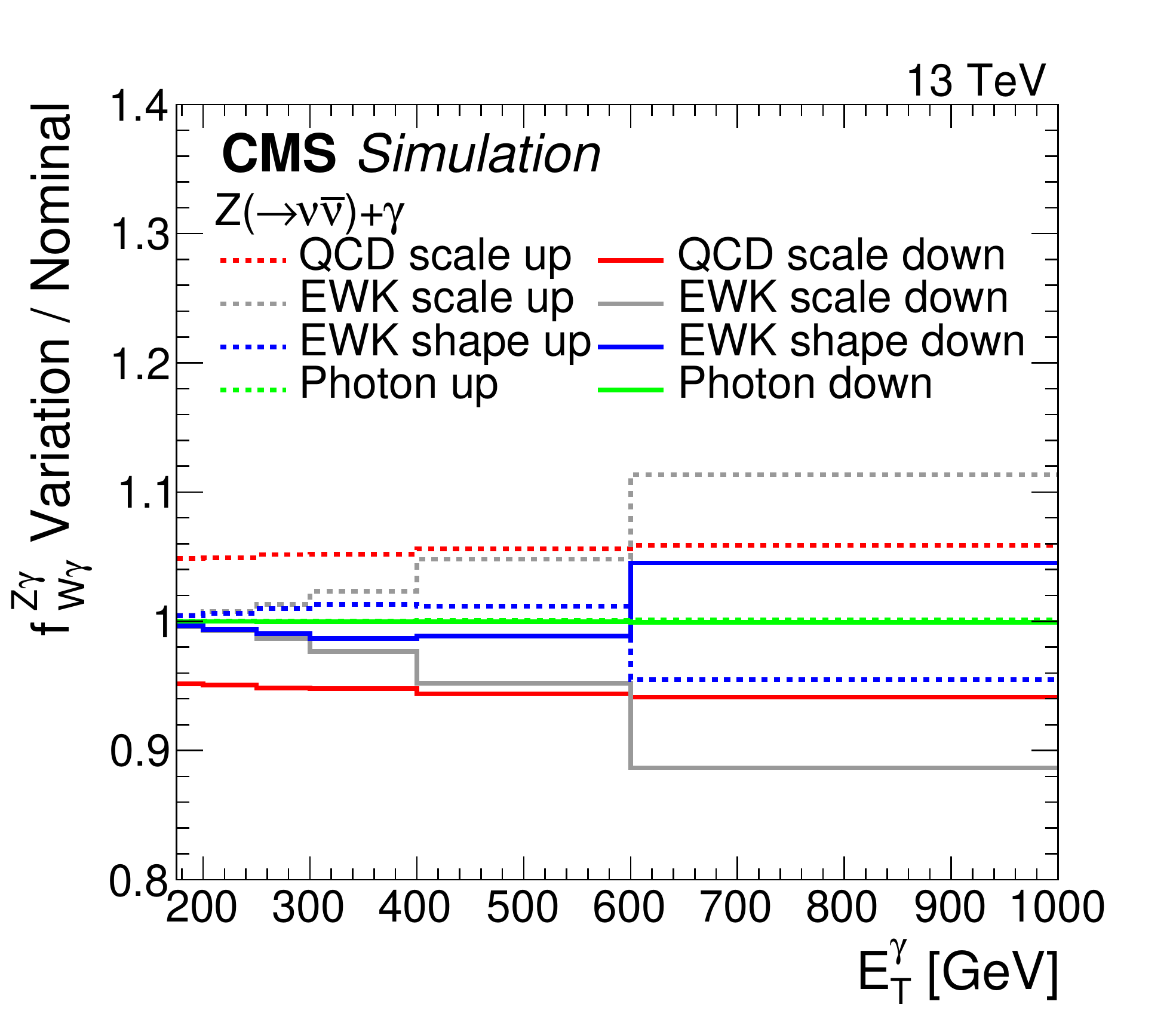}
    \includegraphics[width=0.45\textwidth]{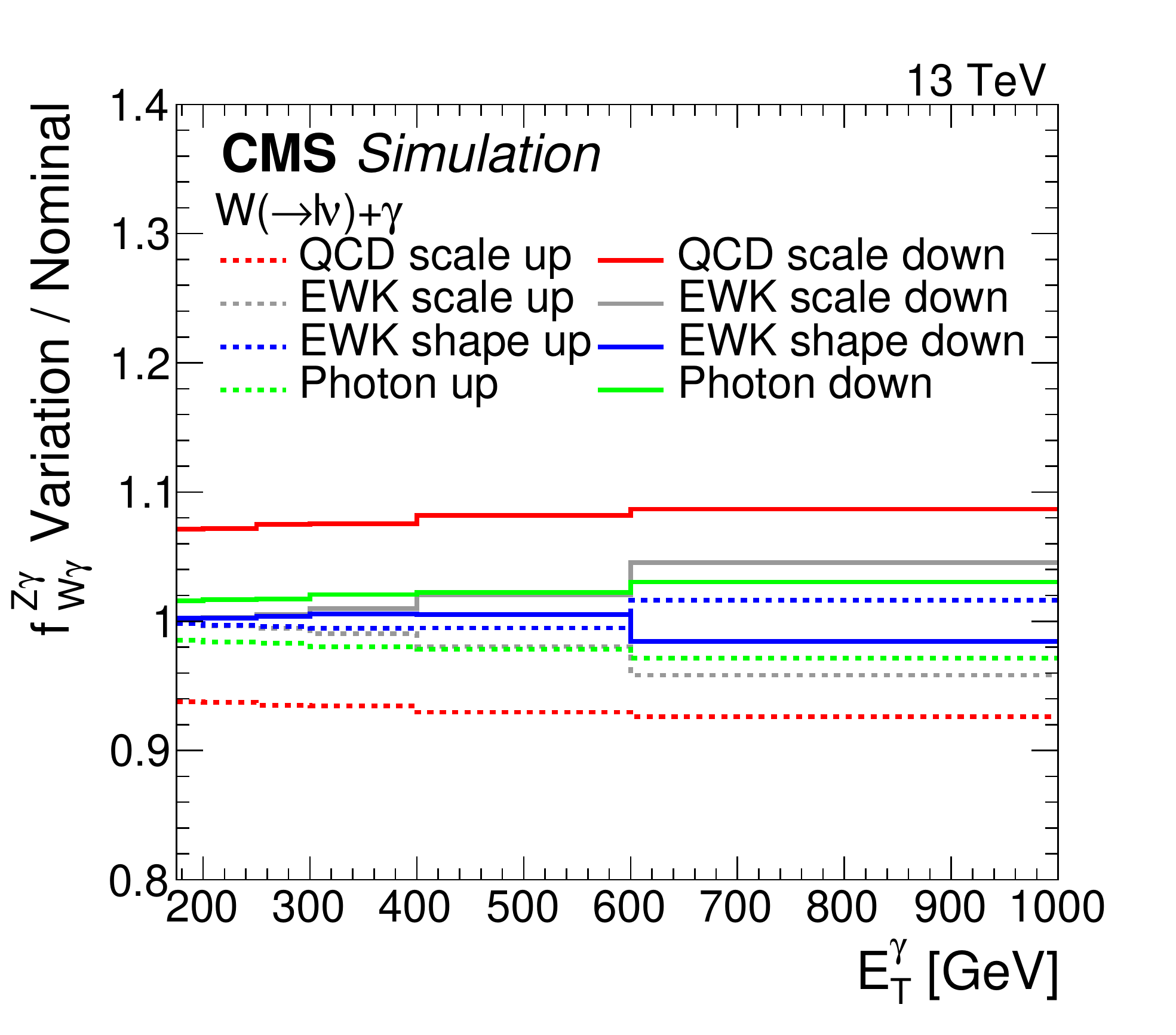}
    \caption{
      Systematic uncertainty in the transfer factors for \zinvg\ (left) and \wlng\ (right). The last bin includes all events with $\ETg > 1000\GeV$.
    }
    \label{fig:tf_syst}
\end{figure}

Figure~\ref{fig:tf_syst} shows the effect of systematic uncertainty in the ratio between the \zinvg\ and \wlng\ processes with respect to nominal value for ${\PZ\Pgg}$ and ${\PW\Pgg}$ respectively.

\section{Simplified Likelihood}
\begin{figure*}[!h]
    \centering
    \includegraphics[width=0.46\textwidth]{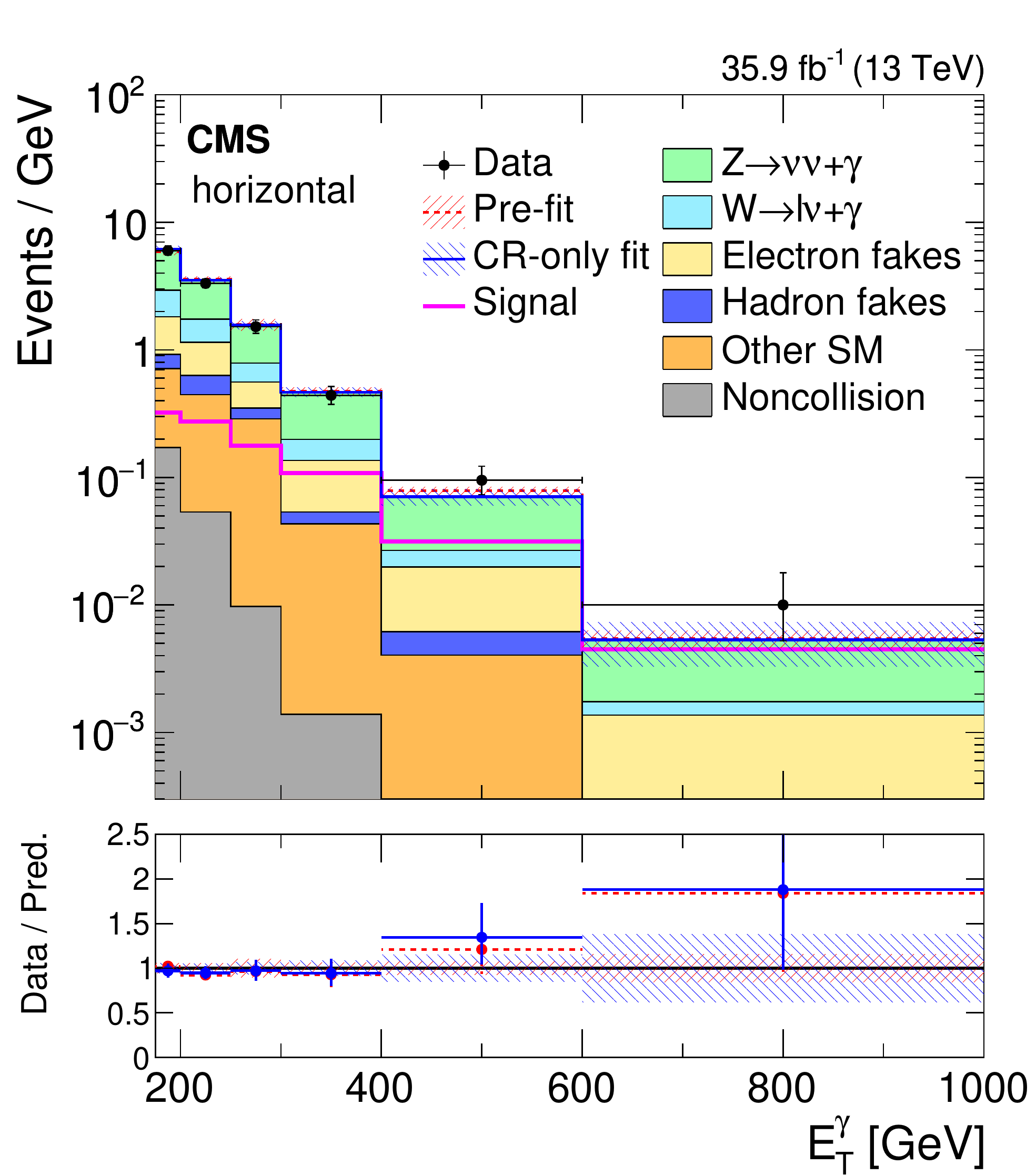}
    \includegraphics[width=0.46\textwidth]{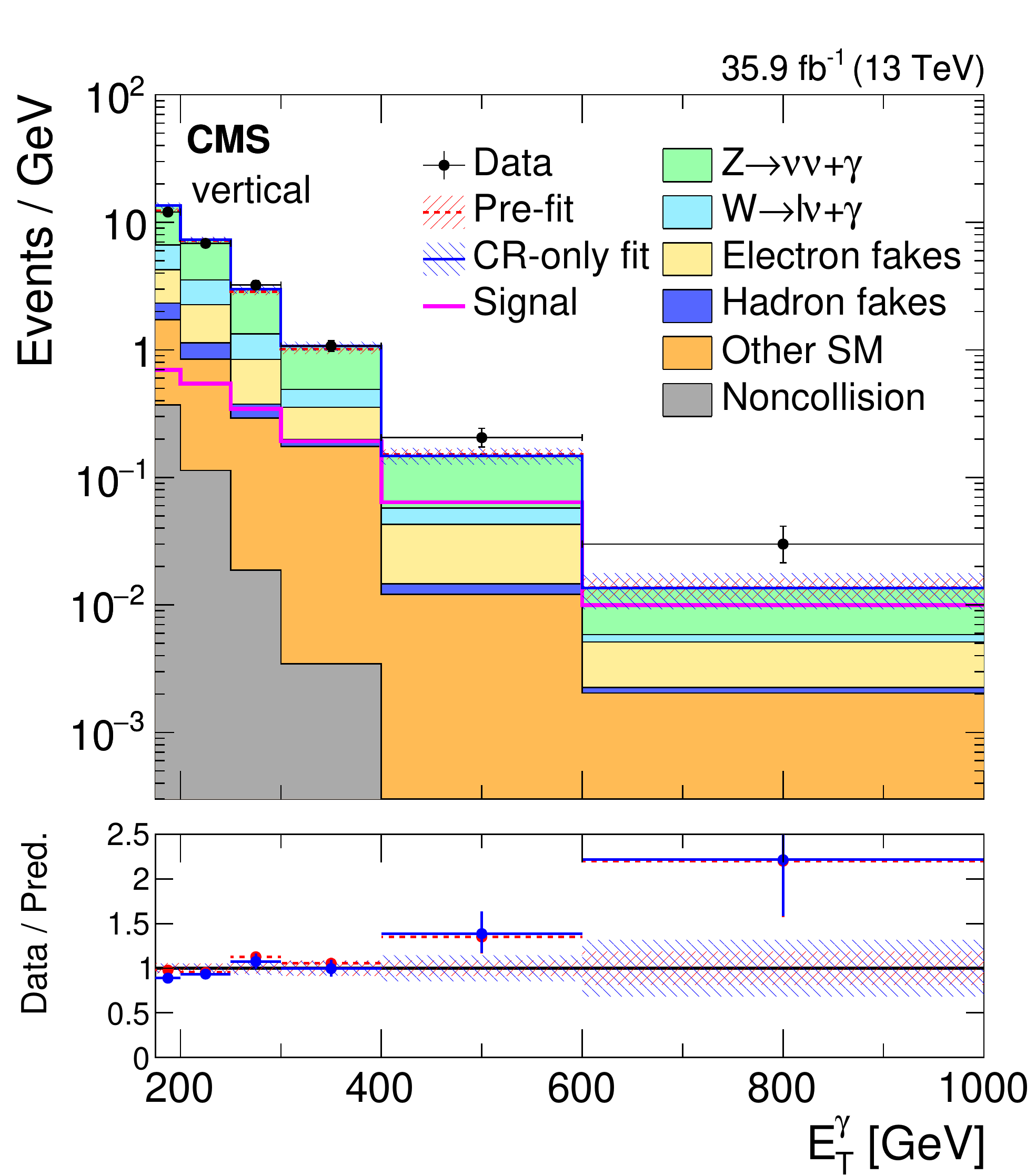}
    \caption{
      Observed \ETg~distribution in the horizontal (left) and vertical (right)
      signal regions compared with the post-fit background expectations for various SM processes.
      The last bin includes all events with $\ETg > 1000\GeV$.
      The expected background distributions are evaluated after performing a
      combined fit to the data in all the control samples, not including the signal region.
    }
    \label{fig:signal_CRonly}
\end{figure*}

\begin{figure*}[hbtp]
 \centering
  \includegraphics[width=0.9\textwidth]{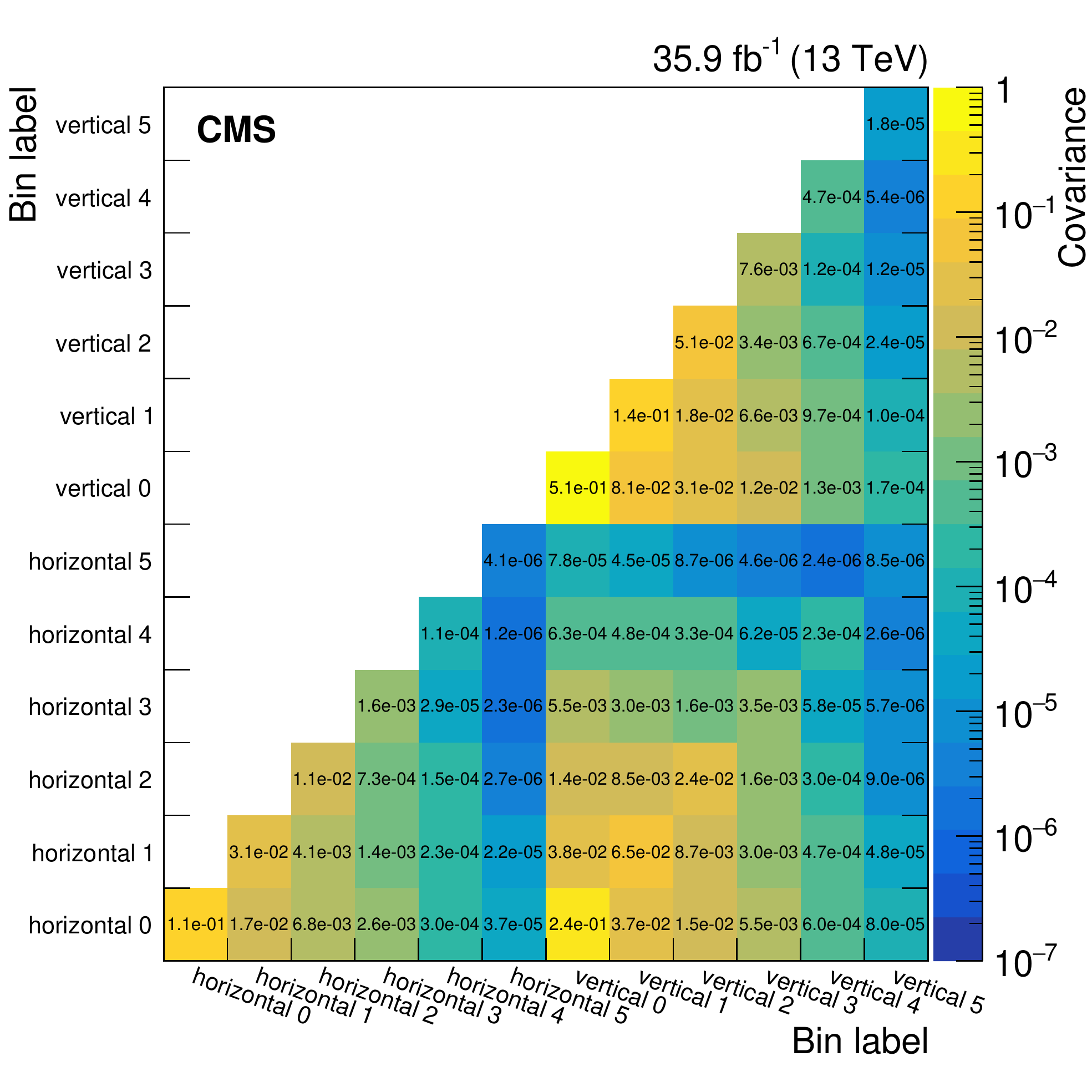}
  \caption{Covariances between the predicted background yields in all the \ETg~bins of the horizontal and vertical signal regions.
           The bin labels specify which signal region the bin belongs to and what number bin it is for that region.}
  \label{fig:correlation_matrix}
\end{figure*}

Figure~\ref{fig:signal_CRonly} shows the comparison between data and the post-fit
background predictions in the horizontal and vertical signal regions, where
the background prediction is obtained from a combined fit performed in all control regions,
excluding the signal regions.  The covariances between the predicted background
yields across all the \ETg~bins in the two signal regions are shown in Fig.~\ref{fig:correlation_matrix}.

Additionally, Table~\ref{tab:cutflow_efficiency} shows the step-by-step efficiency of various selections
for the irreducible $\PZ\Pgg$ and $\PW\Pgg$ processes as well as two representative signal models.

\begin{table}[hbtp]
\topcaption{Step-by-step efficiencies of various selections for irreducible $\PZ\Pgg$ and $\PW\Pgg$ processes as well as two representative signal models:
         a 1\TeV vector mediator decaying to 1\GeV DM particles and an ADD graviton model with 8 extra dimensions and $\mD = 3\TeV$.
         The statistical uncertainties on these values are generally on the order of half a percent.}
\centering
\begin{tabular}{ lcccc }
\hline
                               &         $\PZ\Pgg$ &         $\PW\Pgg$ &         Vector mediator  &         ADD graviton \\
\hline
Trigger                        &         0.4498    &         0.4750    &         0.6348              &         0.6610       \\
Photon selection               &         0.1832    &         0.1960    &         0.3194              &         0.3664       \\
$\ptmiss > 170\GeV$           &         0.1064    &         0.0297    &         0.2800              &         0.3305       \\
Lepton veto                    &         0.1055    &         0.0148    &         0.2781              &         0.3283       \\
$\DphiMETg > 0.5$              &         0.1047    &         0.0134    &         0.2271              &         0.3283       \\
$\minDphiMETj > 0.5$           &         0.0928    &         0.0084    &         0.2512              &         0.3004       \\
$\ETg~/ \ptmiss < 1.4$         &         0.0892    &         0.0074    &         0.2477              &         0.2959       \\
\hline
\end{tabular}
\label{tab:cutflow_efficiency}
\end{table}
\cleardoublepage \section{The CMS Collaboration \label{app:collab}}\begin{sloppypar}\hyphenpenalty=5000\widowpenalty=500\clubpenalty=5000\vskip\cmsinstskip
\textbf{Yerevan Physics Institute, Yerevan, Armenia}\\*[0pt]
A.M.~Sirunyan, A.~Tumasyan
\vskip\cmsinstskip
\textbf{Institut f\"{u}r Hochenergiephysik, Wien, Austria}\\*[0pt]
W.~Adam, F.~Ambrogi, E.~Asilar, T.~Bergauer, J.~Brandstetter, M.~Dragicevic, J.~Er\"{o}, A.~Escalante~Del~Valle, M.~Flechl, R.~Fr\"{u}hwirth\cmsAuthorMark{1}, V.M.~Ghete, J.~Hrubec, M.~Jeitler\cmsAuthorMark{1}, N.~Krammer, I.~Kr\"{a}tschmer, D.~Liko, T.~Madlener, I.~Mikulec, N.~Rad, H.~Rohringer, J.~Schieck\cmsAuthorMark{1}, R.~Sch\"{o}fbeck, M.~Spanring, D.~Spitzbart, A.~Taurok, W.~Waltenberger, J.~Wittmann, C.-E.~Wulz\cmsAuthorMark{1}, M.~Zarucki
% ...
\vskip\cmsinstskip
% ...
1:  Also at Vienna University of Technology, Vienna, Austria\\
% ...
\end{sloppypar}
\end{document}